\documentclass[a4paper,11pt]{article}
\usepackage{lineno}
\usepackage{amssymb}
\usepackage{amsfonts,amsbsy,graphics,epsfig}
\usepackage{graphicx}
\usepackage{amsmath}
\usepackage{color}
\usepackage{multirow}
\usepackage{arydshln}
\newtheorem{theorem}{Theorem}[section]
\newtheorem{lemma}{Lemma}[section]
\newtheorem{proposition}{Proposition}[section]
\newtheorem{corollary}{Corollary}[section]
\newtheorem{remark}{Remark}

\usepackage[dvipsnames]{xcolor}
\usepackage{wasysym}
\usepackage{natbib}
\usepackage{cleveref}
\usepackage{adjustbox}
\modulolinenumbers[1]

\begin{document}
\begin{center}	
	\textbf{\Huge Partially Linear Spatial Probit Models}\medskip \vspace{0.5cm}\\
	\Large{Mohamed-Salem AHMED \\
	{University of Lille, LEM-CNRS 9221\\Lille, France\\ mohamed-salem.ahmed@univ-lille.fr}\\ 
	\vspace{0.5cm}
	
	{Sophie DABO }\\
	INRIA-MODAL\\
	{University of Lille LEM-CNRS 9221\\ 
	Lille, France \\ sophie.dabo@univ-lille.fr}}\\

\end{center}	
\begin{center}
	\rule{1\linewidth}{.9pt}
\end{center}
\noindent	
\textbf{\Large Abstract}\medskip \\
		A partially linear probit model for spatially dependent data is considered. A triangular array setting is used to cover various patterns of spatial data. Conditional spatial heteroscedasticity and non-identically distributed observations and a linear process for disturbances are assumed, allowing various spatial dependencies. The estimation procedure is a combination of a weighted likelihood and a generalized method of moments. The procedure first fixes the parametric components of the model and then estimates the non-parametric part using weighted likelihood; the obtained estimate is then used to construct a GMM parametric component estimate. The consistency and asymptotic distribution of the  estimators are established under sufficient conditions. Some simulation experiments  are provided to investigate the finite sample performance of the estimators.\medskip \\
\textit{keyword:}
		Binary choice model, GMM, non-parametric statistics, spatial econometrics, spatial statistics.
\begin{center}
	\rule{1\linewidth}{.9pt}
\end{center}
 \section*{Introduction}
Agriculture, economics, environmental sciences, urban systems, and epidemiology activities often utilize spatially dependent data. Therefore, modelling such activities requires one to find a type of correlation between some random variables in one location with other  variables in neighbouring locations; see for instance \cite{pinkse1998contracting}. This
is a significant feature of spatial data analysis. Spatial/Econometrics statistics provides tools to perform such modelling. Many studies on spatial effects in statistics and econometrics using many diverse models have been published; see \cite{cressie15}, \cite{anselin2010thirty}, \cite{anselin2013spatial} and \cite{arbia2006spatial} for a review.\\
Two main methods of incorporating  a spatially dependent structure \citep[see for instance][]{cressie15} can essentially be distinguished as between geostatistics and lattice data. In the domain of geostatistics, the spatial location is valued in a continuous set of $\mathbb{R}^N$, $N\ge 2$. 
However, for many activities, the spatial index or location does not vary continuously and may be of the lattice type, the baseline of this current work. In image analysis, remote sensing from satellites, agriculture etc., data are often received as a regular lattice and identified as the centroids of square pixels, whereas a mapping often forms an irregular lattice. Basically, statistical models for lattice data are linked to nearest neighbours to express the fact that data are nearby. \\  Two popular spatial dependence models have received substantial attention for lattice data, the spatial autoregressive (SAR) dependent variable model and the spatial autoregressive error model (SAE, where the model error is an SAR), which extend the regression in a time series to spatial data.\\
From a theoretical point of view, various linear spatial regression SAR and SAE models as well as their identification and estimation methods, e.g., two-stage least squares (2SLS), three-stage least squares (3SLS),  maximum likelihood (ML) or quasi-maximum likelihood (QML) and the generalized method of moments (GMM), have been developed and summarized by many authors such as \cite{anselin2013spatial},  \cite{kelejian1998}, \cite{kelejian1999}, \cite{conley1999gmm}, \cite{cressie15}, \cite{Case},  \cite{lee2004asymptotic}, \cite{lee2007}, \cite{leeetal10}, \cite{zhendetal12}, \cite{malikovetal17}, \cite{garthoffetal17}, \cite{yangetal17}.
Introducing nonlinearity into the field of spatial linear lattice models has attracted less attention; see for instance
\cite{robinson2011asymptotic}, who generalized kernel regression estimation to spatial lattice data. \cite{su2012semiparametric} proposed a semi-parametric GMM estimation for some semi-parametric SAR models. 
Extending these models and methods to discrete choice spatial models has seen less attention; only  a few papers were have been concerned with this topic in recent years. This may be, as noted by \cite{fleming2004techniques} (see also \cite{smirnov2010modeling} and \cite{bille2014computational}), due to the "added complexity that spatial dependence introduces into discrete choice models". Estimating the model parameters with a full ML approach in spatially discrete choice models often requires solving a very computationally demanding problem of $n$-dimensional integration, where $n$ is the sample size.\\  {For linear models, many discrete choice models are fully linear and utilize
	a continuous latent variable; see for instance \cite{smirnov2010modeling}, \cite{wang2013partial} and \cite{martinetti2017approximate}, who proposed pseudo-ML methods, and \cite{pinkse1998contracting}, who studied a method  based on the GMM approach.  Also, others methodologies of estimation are emerged like,  EM algorithm \citep{mcmillen1992probit} and Gibbs sampling approach \citep{lesage2000bayesian}.}

When the relationship between the discrete choice variable and some explanatory variables is not linear, a semi-parametric model may represent an alternative to fully parametric models. This type of model is known in the literature as \textit{partially linear choice spatial models} and is the baseline of this current work. 
When the data are independent, these  choice models can be viewed as special cases of the famous generalized additive models \citep[][]{hastie1990generalized} and have received substantial attention in the literature, and various estimation methods have been explored \citep[see for instance][]{hunsberger1994semiparametric, severini1994quasi, carroll1997generalized}. \\
To the best of our knowledge, semi-parametric spatial choice models have not yet been investigated from a theoretical point of view.  
To fill this gap, this work addresses an SAE spatial probit model for when the spatial dependence structure is integrated in a disturbance term of the studied model.\\
We propose a semi-parametric estimation method combining the GMM approach  and the weighted likelihood method. The method consists of first fixing the parametric components of the model  and  non-parametrically estimating the non-linear component by weighted likelihood \citep{staniswalis1989kernel}.  The obtained estimator depending on the values at which the parametric components are fixed is  used to construct a GMM estimator \citep{pinkse1998contracting} of these components. \\
\noindent The remainder of this paper is organized as follows. In Section \ref{sec1}, we introduce the studied spatial model and the estimation procedure. Section~\ref{sec3} is devoted to  hypotheses and asymptotic results, while Section~\ref{sec4} reports a discussion and computation of the estimates. Section~\ref{sec5} gives some numerical results based on simulated data to illustrate the performance of the proposed
estimators. The last section presents the proofs of the main results. 

\section{Model}\label{sec1}
We consider that at $n$ spatial locations $\{s_1,s_2,\ldots,s_n\}$ satisfying $\Vert s_i-s_j\Vert > \rho $ with $\rho>0 $,  observations of a random vector  $(Y,X,Z)$ are available. Assume that these observations are considered as  triangular arrays \citep{robinson2011asymptotic} and follow the partially linear model of a  latent dependent variable $Y^{*}$:   
\begin{equation}
Y^{*}_{in}=X_{in}^{T}\beta_0 + g_0(Z_{in})+U_{in},\qquad 1\leq i\leq n,\; n=1,2,\ldots
\label{md1}
\end{equation}
with 
\begin{equation}
Y_{in}=\mathbb{I}\left(Y^{*}_{in}\geq 0\right) ,\qquad 1\leq i\leq n,\; n=1,2,\ldots
\label{md2}
\end{equation} 
where $\mathbb{I}(\cdot)$ is the indicator function; $X$ and $Z$ are explanatory random variables taking values in the two compact subsets $\mathcal{X}\subset \mathbb{R}^{p} (p\geq 1)$ and $\mathcal{Z}\subset \mathbb{R}^{d} (d\geq 1)$, respectively; the parameter $\beta_0$ is an unknown $p\times 1$ vector that belongs to a compact subset $\Theta_\beta \subset \mathbb{R}^{p}$;  and $g_0(\cdot)$ is an unknown smooth function valued in the space of functions $\mathcal{G}=\left \{g\in C^{2}(\mathcal{Z}): \Vert g\Vert=\sup_{z\in\mathcal{Z}}\vert g(z)\vert <C \right\}$, with $C^{2}(\mathcal{Z})$ the space of twice differentiable functions from $\mathcal{Z}$ to $\mathbb{R}$ and $C$ a positive constant. In model (\ref{md1}), $\beta_0$ and $g_0(\cdot)$ are constant over $i$ (and $n$). Assume that the disturbance term $U_{in} $ in $(\ref{md2})$  is modelled by the following  spatial autoregressive process (SAR):
\begin{equation}
U_{in}=\lambda_0\sum_{j=1}^{n}W_{ijn}U_{jn}+\varepsilon_{in}, \qquad 1\leq i\leq n,\; n=1,2,\ldots
\label{SAR1}
\end{equation}     
where $\lambda_0$ is the autoregressive parameter, valued in the compact subset $\Theta_{\lambda}\subset \mathbb{R}$, $W_{ijn},\, j=1,...,n$ are the elements in the $i$--th row of a non-stochastic  $n\times n $ spatial weight matrix $W_{n}$, which contains the information on the spatial relationship between observations. This spatial weight matrix is usually constructed as a function of the distances (with respect to some metric) between locations; see \cite{pinkse1998contracting} for additional details. 
The $n\times n$ matrix $(I_n- \lambda_0 W_n)$ is assumed to be  non-singular for all $n$, where $I_n$ denotes  the $n\times n$ identity matrix and $\left\{\varepsilon_{in}, \; 1\leq i \leq n\right\}$ are assumed to be independent random Gaussian variables; $\mathbb{E}(\varepsilon_{in})=0$ and $\mathbb{E}(\varepsilon_{in}^2)=1$ for $i=1,\ldots,n\; n=1,2,\ldots$. Note that one can rewrite $(\ref{SAR1})$ as 

\begin{equation}
U_{n}=\left(I_n -\lambda_0 W_n\right)^{-1}\varepsilon_n, \quad n=1,2,\ldots
\end{equation}
where $U_n=(U_{n1},\ldots,U_{nn})^{T}$ and $\varepsilon_n=(\varepsilon_{n1},\ldots,\varepsilon_{nn})^{T}$. Therefore, the variance-covariance matrix of $U_n$ is 
\begin{equation}
V_{n}(\lambda_0)\equiv \mathrm{Var}(U_n)=\left(I_n -\lambda_0 W_n\right)^{-1}\left\{\left(I_n -\lambda_0 W_n\right)^{T}\right\}^{-1},\quad n=1,2,\ldots
\end{equation}

This matrix  allows one to describe the cross-sectional spatial dependencies between the $n$ observations. Furthermore,  the fact that the diagonal elements of $V_{n}(\lambda_0)$ depend on $\lambda_0$ and particularly  on $i$ and $n$   allows some spatial heteroscedasticity. These spatial dependences and heteroscedasticity depend on the neighbourhood structure established by the spatial weight matrix $W_n$.
\\
Before proceeding further, let us give some particular cases of the model.\\
If one consider i.i.d observations, that is, $V_n(\lambda_0)=\sigma^2I_n,$ with $\sigma$ depending on $\lambda_0$, the obtained model
may be viewed as a special case of classical generalized partially linear models \citep[e.g.][]{severini1994quasi} or the classical generalized additive model \citep{hastie1990generalized}. Several approaches for estimating this particular model have been developed; among these methods, we cite that of \cite{severini1994quasi} based on the concept of the generalized profile likelihood \cite[e.g][]{severini1992profile}. This approach consists of first fixing the parametric parameter $\beta$ and  non-parametrically estimating  $g_0(\cdot)$ using the weighted likelihood method. This last estimate is  then used to construct a profile likelihood  to estimate $\beta_0$.\\
When $g_0\equiv 0$ (or is an affine function), that is,  without a non-parametric component, several approaches have been developed to estimate the parameters $\beta_0$ and $\lambda_0$.  The basic difficulty encountered is that the likelihood function of this model involves an $n$-dimensional normal integral; thus, when $n$ is high, the computation or asymptotic properties of the estimates may present difficulties  \citep[e.g.][]{poirier1988probit}. Various approaches have been proposed to addressed this difficulty; among these approaches, we cite the following:
\begin{itemize}
	\item Feasible Maximum Likelihood approach: this approach consists  of replacing the true likelihood function by a pseudo-likelihood function constructed via marginal likelihood functions. \cite{smirnov2010modeling} proposed  a pseudo-likelihood function obtained by  replacing  $V_{n}(\lambda_0)$  by some diagonal matrix with  the diagonal elements of $V_n(\lambda_0)$. Alternatively, \cite{wang2013partial} proposed to  divide the observations by pairwise groups, where the latter  are  assumed to be independent with a bivariate normal distribution in each group, and estimate  $\beta_0$ and $\lambda_0$ by maximizing the likelihood of these groups. {Recently \cite{martinetti2017approximate}
		proposed a pseudo-likelihood function defined as an approximation of the likelihood function where the latter is inspired by  some univariate conditioning procedure.}
	\item Generalized Method of Moments (GMM) approach used by \cite{pinkse1998contracting}. These authors  used the generalized residuals defined by  $\tilde{U}_{in}(\beta,\lambda)=\mathbb{E}\left(U_{in}\vert Y_{in},\beta,\lambda\right), \; 1\leq i\leq n, \; n=1,2,\ldots$
	with some instrumental variables to construct  moment equations to define the GMM estimators of $\beta_0$ and $\lambda_0$.
\end{itemize} 
In what follows, using the $n$ observations $(X_{in}, Y_{in}, Z_{in}),\, i=1,...,n$, we  propose parametric estimators of $\beta_0$, $\lambda_0$ and  a non-parametric estimator of the smooth function $g_0(\cdot)$.\\
To this end, we assume that, for all $n=1,2,\ldots$, $\left\{\varepsilon_{in}, \; 1\leq i \leq n\right\}$ is independent of $\left\{X_{in}, \; 1\leq i \leq n\right\}$ and $\left\{Z_{in}, \; 1\leq i \leq n\right\}$, and $\left\{X_{in}, \; 1\leq i \leq n\right\}$ is independent of  $\left\{Z_{in}, \; 1\leq i \leq n\right\}$.\\ 
We give asymptotic results according to \textit{increasing domain} asymptotic. This consists of a sampling structure whereby new observations are added at the edges (boundary points) to compare to the \textit{infill} asymptotic, which consists of a sampling structure whereby  new  observations are added in-between existing observations. A typical example of an increasing domain is lattice data. An infill asymptotic is appropriate when the spatial locations are in a bounded domain. 

\subsection{Estimation Procedure}\label{sec2}

We propose an estimation procedure based on  a combination of a weighted likelihood method and a generalized method of moments. We first fix the parametric components $\beta $ and $\lambda$ of the model and estimate the non-parametric component  using a weighted likelihood. The obtained estimate is then used to construct generalized residuals, where  the latter are combined with the instrumental variables  to propose GMM parametric estimates. This approach will be described as follow.  	     

By equation (\ref{md2}), we have  
\begin{equation}
\mathbb{E}_0\left(Y_{in}\vert X_{in},Z_{in}\right)=\Phi\left(\left( v_{in}(\lambda_0)\right)^{-1}\left(X_{in}^{T}\beta_0+g_0(Z_{in})\right)\right), \quad 1\leq i\leq n, \quad n=1,2,\ldots
\label{md3}
\end{equation}
where $\mathbb{E}_0$ denotes the expectation under the true parameters (i.e.,  $\beta_0, \lambda_0$ and $g_0(\cdot)$), $\Phi(\cdot)$ is the cumulative distribution function of a standard  normal distribution,  and  $(v_{in}(\lambda_0))^2= V_{iin}(\lambda_0),\; 1\leq i\leq n, \; n=1,2,\cdots $ are the diagonal elements of $V_{n}(\lambda_0)$. \\ For each $\beta \in \Theta_\beta$, $\lambda \in \Theta_\lambda, \, z \in \mathcal{Z}$ and  $\eta \in \mathbb{R}$, we define the conditional expectation on  $Z_{in}$ of the log-likelihood of $Y_{in}$ given $(X_{in},Z_{in}) $  for $ 1\leq i \leq n, \; n=1,2,\ldots$, as
\begin{equation}
H(\eta; \beta, \lambda, z)=\mathbb{E}_{0}\left(\left.\mathcal{L}\left(\Phi\left(\left( v_{in}(\lambda)\right)^{-1}\left(\eta + X_{in}^{T}\beta\right)\right);Y_{in} \right)\right\vert Z_{in}=z\right),
\end{equation} 
with $\mathcal{L}(u;v)=\log \left(u^{v}(1-u)^{1-v}\right)$. Note that $H(\eta; \beta, \lambda, z)$ is assumed to be constant over $i$ (and $n$).  For each fixed $\beta \in \Theta_\beta$, $\lambda \in \Theta_\lambda$ and $z \in \mathcal{Z}$,  $g_{\beta,\lambda}(z)$ denotes the solution in $\eta$ of
\begin{equation}
\frac{\partial}{\partial\eta}H(\eta; \beta, \lambda, z)=0.
\label{partialH}
\end{equation}
Then, we have $g_{\beta_{0},\lambda_{0}}(z)=g_{0}(z) $ for all $z\in \mathcal{Z}$.\\
Now, using $g_{\beta,\lambda}(\cdot)$, we construct the GMM estimates  of $\beta_0$ and $\lambda_0$ as in \cite{pinkse1998contracting}. For that, we define the generalized residuals, replacing $g_0(Z_{in})$ in $(\ref{md1})$ by $g_{\beta,\lambda}(Z_{in})$: 
\begin{eqnarray}
\tilde{U}_{in}(\beta,\lambda,g_{\beta,\lambda})&=&\mathbb{E}\left(U_{in}\vert Y_{in},\beta,\lambda\right) \label{Resud}\\
&=&\frac{\phi\left(G_{in}(\beta,\lambda,g_{\beta,\lambda})\right)\left(Y_{in}-\Phi\left(G_{in}(\beta,\lambda,g_{\beta,\lambda})\right)\right)}{\Phi\left(G_{in}(\beta,\lambda, g_{\beta,\lambda})\right)\left(1-\Phi\left(G_{in}(\beta,\lambda, g_{\beta,\lambda})\right)\right)},\nonumber 
\end{eqnarray}
where $\phi(\cdot)$ is the density of the standard normal distribution and\\
$G_{in}(\beta,\lambda, g_{\beta,\lambda})=\left( v_{ni}(\lambda)\right)^{-1}\left(X_{in}^{T}\beta+ g_{\beta,\lambda}(Z_{in})\right).$\\
For simplicity of notation, we write   $\theta=(\beta^{T},\lambda)^{T} \in \Theta\equiv \Theta_\beta \times \Theta_\lambda$ when  possible.\\
Note that in (\ref{Resud}), the generalized residual $\tilde{U}_{in}(\cdot\, ,\, \cdot)$ is calculated  by conditioning only on $Y_{in}$ and not on the entire sample $\{Y_{in},\, i=1,2,\ldots,n,\, n=1,\ldots\}$ or a subset of it. This of course will influence the efficiency of the estimators of $\theta$ obtained by these generalized residuals, but it allows one to avoid  a complex computation; see \cite{poirier1988probit} for additional details. To address this loss of efficiency, let us follow \cite{pinkse1998contracting}'s procedure, which consists of employing some instrumental variables to create some moment conditions, and use a random matrix to define a criterion function. Both the instrumental variables and the random matrix permit one to consider more information about the spatial dependences and heteroscedasticity characterizing the dataset. Let us now detail the estimation procedure.
Let 
\begin{equation}
S_{n}(\theta, g_{\theta})=n^{-1}\xi_n^{T}\tilde{U}_n(\theta,g_{\theta}),
\label{eq0}
\end{equation}
where $\tilde{U}_n(\theta,g_{\theta})$ is an $n\times 1$ vector, composed of $\tilde{U}_{in}(\theta,g_{\theta}), \; 1\leq i\leq n$ and $\xi_n$ is an $n\times q$ matrix of instrumental variables, whose $i$th row is  given by the $1\times q$ random vector $\xi_{in}$. The latter  may depend on $g_{\theta}(\cdot)$ and $\theta$. 
We assume that $\xi_{in}$ is $\sigma(X_{in},Z_{in})$, measurable for each $i=1,\ldots,n, \; n=1,2,\ldots$.  
We suppress the possible dependence of the instrumental variables on the  parameters for notational simplicity. 
The GMM approach consists of minimizing the following sample criterion  function:
\begin{equation}
Q_{n}(\theta, g_{\theta})=S_{n}^{T}(\theta, g_{\theta})M_{n}S_{n}(\theta, g_{\theta}),
\label{gmm}
\end{equation}
where $M_{n}$ is some positive-definite $q\times q$ weight matrix that may depend on the sample information. The choice of the instrumental variables and  weight matrix characterizes the difference between GMM estimator  and all pseudo-maximum likelihood estimators. For instance, if one takes  
\begin{equation}
\xi_{in}(\theta,g_{\theta})=\frac{\partial G_{in}(\theta,\eta_i)}{\partial\theta}+\frac{\partial G_{in}(\theta,\eta_i)}{\partial\eta}\, \frac{\partial g_{\theta}}{\partial \theta}(Z_{in}),
\label{CIV}
\end{equation}
with $\eta_i=g_{\theta}(Z_{in})$, $G_{in}(\theta, \eta_i)=\left( v_{in}(\lambda)\right)^{-1}\left(X_{in}^{T}\beta+ \eta_i\right)$, and $M_n=I_{q}$ with  $q=p+1$, then  the GMM estimator of $\theta$ is equal to a pseudo-maximum profile likelihood estimator of  $\theta$, accounting  only for the spatial heteroscedasticity.\\
Now, let
\begin{equation}
S(\theta, g_{\theta})=\lim_{n\rightarrow\infty}\mathbb{E}_{0}\left(S_{n}(\theta, g_{\theta})\right),
\label{DefS}
\end{equation}
and $$Q(\theta,g_{\theta})=S^{T}(\theta, g_{\theta})MS(\theta, g_{\theta}),$$
where $M$, the limit of the sequence $M_{n}$,  is a nonrandom positive-definite matrix. The functions  $S_n(\cdot,\cdot)$ and $Q_n(\cdot,\cdot)$ are viewed as empirical counterparts of $S(\cdot,\cdot)$ and $Q(\cdot,\cdot)$, respectively. \\
Clearly,  $g_{\theta}(\cdot) $ is not available in practice. However, we need to estimate it, particularly by an asymptotically efficient estimate. By ($\ref{partialH}$) and for fixed  $\theta^{T}=(\beta^T,\lambda) \in \Theta$, an estimator of  $g_{\theta}(z)$, for $z \in \mathcal{Z}$, can be given by $\hat{g}_{\theta}(z)$, which  denotes the solution in $\eta$ of 
\begin{equation}
\sum_{i=1}^{n}\frac{\partial}{\partial\eta} \mathcal{L}
\left(\Phi\left(G_{in}(\theta, \eta)\right);Y_{in}\right)K\left(\frac{z-Z_{in}}{b_n}\right)=0
\label{eq1}
\end{equation}
where   $K(\cdot)$ is a kernel from $\mathbb{R}^{d}$ to $\mathbb{R}_{+}$ and $b_n$ is a  bandwidth depending  on ${n}$.\\

Now,  replacing $g_{\theta}(\cdot)$ in $(\ref{gmm})$  by the  estimator $\hat{g}_{\theta}(\cdot)$ permits one to  obtain the GMM estimator $\hat{\theta}$ of $\theta$ as
\begin{equation}
\hat{\theta}=\mathrm{argmin}_{\theta\in \Theta}Q_{n}(\theta,\hat{g}_{\theta}).
\label{eq2}
\end{equation}
A classical inconvenience  of the estimator $\hat{g}_{\theta}(z)$ proposed in (\ref{eq1}) is that the bias of $\hat{g}_{\theta}(z)$ is high for $z$ near the boundary of $\mathcal{Z}$. Of course, this bias will affect the estimator of $\theta$ given in (\ref{eq2}) when some of the observations $Z_{in}$ are near  the boundary of  $\mathcal{Z}$. A local linear method, or more generally the local polynomial  method \citep{fan1996local}, can be used to reduce this bias. Another alternative is to use \textit{trimming} \citep{severini1994quasi}, in which the function $S_n(\theta,g_{\theta})$ is computed using only observations associated with $Z_{in}$ that are away from the boundary. The advantage of this approach is that the theoretical results can be presented in a clear form, but it is less tractable from a practical point of view, in particular, for small sample sizes. 
\section{Large sample properties}\label{sec3}

We now turn to the asymptotic properties of the estimators derived in the previous section: $\hat{\theta}^T=(\hat{\beta}^T,\hat{\lambda})$ and $\hat{g}_{\hat{\theta}}(\cdot)$. 
Let us use the following  notation:
$\frac{d}{d\theta} S(\theta,g_{\theta})$ means that we differentiate $S(.,.)$ with respect to $\theta$, and  $\frac{\partial }{\partial \theta} S(\theta,g_{\theta})$ is the partial derivative of $S(\cdot, \cdot)$ w.r.t the first variable. The partial derivative of  $S_{n}(\theta,g)$ w.r.t $g$, for any function $v\in \mathcal{G}$, is
\begin{equation*}\frac{\partial S_{n}}{\partial g}(\theta,g)(v)=n^{-1}\sum_{i=1}^{n}\xi_{in}\frac{\partial\tilde{U}_{in}}{\partial \eta}(\theta,\eta_{i})v(Z_{in}).
\end{equation*}
Without ambiguity, $\Vert a \Vert$ denotes $\sup_{t}\vert a(t)\vert $ when $a$ is a function, $\left(\sum a_i^2\right)^{1/2}$ when $a$ is a vector, and $\left(\sum\sum a_{ij}^2\right)^{1/2}$ when $a$ is  a matrix. \\
Let the following matrices be needed in the asymptotic  variance-covariance matrix of $\hat{\theta}$:
$$B_{1}(\theta_{0})=\lim_{n\rightarrow \infty} \mathbb{E}_{0}\left(nS_{n}\left(\theta_{0},g_{0}\right)S^{T}_{n}\left(\theta_{0},g_{0}\right) \right), \; B_{2}(\theta_{0})=\left\{\left.\frac{d }{d \theta}S^{T}\left(\theta,g_{\theta}\right)\right\vert_{\theta=\theta_{0}}\right \}M\left\{\left.\frac{d }{d \theta}S\left(\theta,g_{\theta}\right)\right\vert_{\theta=\theta_{0}}\right\},$$
with 
\begin{equation}
\frac{d }{d \theta}S\left(\theta,g_{\theta}\right)=\frac{\partial S }{\partial\theta}\left(\theta,g_{\theta}\right)+\frac{\partial S }{\partial g}\left(\theta,g_{\theta}\right)\frac{\partial}{\partial\theta}g_{\theta},
\label{B20}
\end{equation}
and 
$$\Omega(\theta_0)=\left\{ B_{2}(\theta_{0})\right\}^{-1}\left\{\left.\frac{d  }{d \theta}S^{T}\left(\theta,g_{\theta}\right)\right\vert_{\theta=\theta_{0}}\right \}M B_{1}(\theta_{0})M \left\{\left.\frac{d }{d \theta}S\left(\theta,g_{\theta}\right)\right\vert_{\theta=\theta_{0}}\right\} \left\{ B_{2}(\theta_{0})\right\}^{-1}.$$
The following  assumptions are required to establish the asymptotic results. \\
\medskip
{\bf Assumption A1. (Smoothing condition)}. For each fixed $\theta\in \Theta$ and $z\in \mathcal{Z}$, let $g_{\theta}(z)$ denote the unique solution with respect to  $\eta$ of 
$$ \frac{\partial}{\partial \eta}H(\eta;\theta, z)=0.$$ 
For any $\varepsilon > 0$ and $ g\in \mathcal{G}$, there exists  $\gamma > 0$ such that 
\begin{equation}
\sup_{\theta\in \Theta, z\in \mathcal{Z}} \left \vert \frac{\partial}{\partial \eta}H(  g(z);\theta, z)\right\vert\leq\gamma
\qquad \Longrightarrow  \qquad 
\sup_{\theta\in \Theta, z\in \mathcal{Z}}\left \vert g(z)-g_{\theta}(z) \right \vert\leq \varepsilon. 
\label{RGC}
\end{equation}
{\bf Assumption A2. (Local dependence).} The  density $f_{in}(\cdot)$ of $Z_{in}$  exists,  is  continuous on $\mathcal{Z}$ uniformly on $i$ and $n$ and satisfies   
\begin{equation} \liminf_{n\to \infty} \,\inf_{z\in\mathcal{Z}} \frac{1}{n}\sum_{i=1}^{n}f_{in}(z) \,>0.
\label{barf}
\end{equation}
The joint probability density $f_{ijn}(.,.)$ of $(Z_{in}, Z_{jn})$ exists and is bounded on $\mathcal{Z}\times\mathcal{Z}$  uniformly on $i\neq j$ and $n$. \\
\noindent
{\bf Assumption A3. (Spatial dependence).}	Let $h_{in}^{\theta,\, \eta_i}(\cdot | \cdot,  \cdot)$ denote the conditional log likelihood function of $Y_{in}$ given $(X_{in},Z_{in})$, where $ \eta_i=g(Z_{in})$. Let  $T_{in}$ be the vector $(Y_{in},X_{in},Z_{in})$,  $i=1,\ldots,n\,,\;n=1,2\ldots $, $\tilde{p}=p+1$, and assume that for all  $i,\,l=1,\ldots,n,$ 
\begin{equation}
\left \vert\mathrm{Cov}_0\left(\psi(T_{in}),\psi(T_{ln})\right)\right\vert \leq\left\{\mathrm{Var}_0\left(\psi(T_{in})\right)\mathrm{Var}_0\left(\psi(T_{ln})\right)\right\}^{1/2}\alpha_{iln},
\label{dependence1}
\end{equation}
with 
$$ \psi(T_{in})=K\left(\frac{z-Z_{in}}{b_n}\right) \;\mbox{ or }\; \psi(T_{in})=K\left(\frac{z-Z_{in}}{b_n}\right)\frac{\partial^{j_1+\cdots+j_{\tilde{p}}+r} }{\partial\theta_1^{j_1}\cdots\partial \theta_{\tilde{p}}^{j_{\tilde{p}}}\partial \eta^{r}}h_{in}^{\theta,\, \eta}(Y_{in}|\,X_{in},\,Z_{in}=z),$$
for all $z\in \mathcal{Z},\; \theta \in \Theta, \eta=g(z)$ with $g\in \mathcal{G}  $, and for all nonnegative integers $j_1,\ldots, j_{\tilde{p}}=0,1,2$ and $r=0,\ldots,4,$ such that $j_1+\cdots+j_{\tilde{p}}+r\leq 6$. \\
We assume that
\begin{equation}
\left \vert \mbox{Cov}_0\left( \xi_{itn}\tilde{U}_{in}(\theta,g_{\theta}),\xi_{jsn}\tilde{U}_{jn}(\theta,g_{\theta})\right)\right \vert \leq \left\{ \mbox{Var}_0\left( \xi_{itn}\tilde{U}_{in}(\theta,g_{\theta})\right) \mbox{Var}_0\left( \xi_{jsn}\tilde{U}_{jn}(\theta,g_{\theta})\right) \right\}^{1/2}\alpha_{ijn},
\label{DepH}
\end{equation}
for all $\theta \in \Theta$, $i,j=1,\ldots,n,\; n=1,2,\ldots$ and for any $s,t =1,\ldots,q$,\\ and 
\begin{equation}
\left \vert \mbox{Cov}_0\left( \xi_{in}^{(2)}(\theta_0,\eta^{0}_i),\xi_{jn}^{(2)}(\theta_0,\eta^{0}_j)\right)\right \vert \leq \left\{ \mbox{Var}_0\left( \xi_{in}^{(2)}(\theta_0,\eta^{0}_i)\right) \mbox{Var}_0\left( \xi_{jn}^{(2)}(\theta_0,\eta^{0}_j)\right) \right\}^{1/2}\alpha_{ijn},
\label{DepH2}
\end{equation}
with	
$$\xi_{in}^{(2)}(\theta_0,\eta_i^0):= w^T\xi_i\Lambda\left(G_{in}(\theta_0,\eta_i^0)\right)\phi\left(G_{in}(\theta_0,\eta_i^0)\right)\frac{\partial G_{in} }{\partial \theta}(\theta_0,\eta_i^0),$$
where $\eta_i^0=g_0(Z_i)$  for  each $w\in \mathbb{R}^{q}$ such that $\Vert w\Vert=1$. 

In addition, assume that there is a decreasing (to $0$) positive function $\varphi(\cdot)$ such that $\alpha_{ijn}=O\left(\varphi\left(\Vert s_i-s_j\Vert \right)\right)$, $r^2\varphi(rr^*)/\varphi(r^*)=o(1)$, as $r\to 0$, for all fixed $r^*>0$, where $s_i $ and $s_j$ are spatial coordinates associated with observations $i$ and $j$, respectively. \\
\noindent
\medskip
{\bf Assumption A4. } The kernel $K$ satisfies  $\int K(u)du=1$. It  is Lipschitzian, i.e., there is a positive  constant $C$ such that 
$$\vert K(u)-K(v)\vert \leq C\Vert u-v \Vert\qquad \mbox{for all }\;u,v\in \mathbb{R}^d.$$	
\noindent
\medskip
{\bf Assumption A5.} The bandwidth $b_n$ satisfies $b_n\to 0$ and $nb_n^{3d+1}\to \infty $ as $n\to \infty$.	\\
\noindent
{\bf Assumption A6. } The instrumental variables satisfy $\sup_{ i,\,  n}\Vert \xi_{in}\Vert=O_p(1)$,  where $\xi_{in}$ is the i-th column of the $n\times q$ matrix of instrumental variables $\xi_n$.\\
\noindent
\medskip
{\bf Assumption A7. } $\theta^{T}=(\beta^{T},\lambda)$ takes values in a compact and convex set $\Theta=\Theta_\beta\times \Theta_\lambda \subset \mathbb{R}^{p}\times\mathbb{R}$, and $\theta_{0}^{T}=(\beta_{0}^{T},\lambda_{0})$ is in the interior of $\Theta$.\\
\noindent
\medskip
{\bf  Assumption A8.} $S(\cdot,\cdot )$ is continuous on  both arguments $\theta$ and $g$, and  $Q(\cdot,g_{.})$ attains a unique minimum over $\Theta$ at $\theta_{0}$.\\
\noindent
\medskip
{\bf Assumption A9.}  The square root of the diagonal elements of $V_n(\lambda)$ are twice continuous differentiable functions with respect to $\lambda$ and  
$\displaystyle \sup_{\lambda\in \Theta_\lambda}\left \vert v_{in}^{-1}(\lambda)+\frac{d}{d\lambda}v_{in}(\lambda)+ \frac{d^2}{d\lambda^2}v_{in}(\lambda)\right\vert < \infty $ uniformly on $i$ and $n$.\\
\noindent
\medskip
{\bf Assumption A10. }  $B_{1}(\theta_{0})$ and $B_{2}(\theta_{0})$ are positive-definite matrices, and  $M_n-M=o_p(1)$. 
\noindent
\begin{remark}
	Assumption  A1 ensures the smoothness of $H(.;.,.)$ around its extrema point $g_\theta (.)$; see \cite{severini1994quasi}.  Assumption  A2 is a decay of the local independence condition of the covariates $Z_{in}$, meaning that these variables are not identically distributed; a similar condition can be find in \cite{robinson2011asymptotic}. Condition (\ref{barf}) generalizes  the classical assumption $\inf_z f(z)>0$ used in the case of  estimating the density function $f(\cdot)$ with identically distributed or stationary random variables. This assumption  has been used in  \cite{robinson2011asymptotic} (\textbf{Assumption A7}(x), p. 8). 
	Assumption A3 describes the spatial dependence structure. The processes that we use are not assumed stationary; this  allows for greater generalizability and the dependence structure to change with the sample size $n$ (see \cite{pinkse1998contracting} for more discussion). Conditions (\ref{dependence1}), (\ref{DepH})  and (\ref{DepH2}) are not restrictive. When the regressors and instrumental variables are deterministic,  conditions (\ref{dependence1})  and (\ref{DepH}) are equivalent to  $\left \vert \mbox{Cov}_0(Y_{in},Y_{ln})\right \vert \leq \alpha_{iln}$. The condition on $\varphi(\cdot)$ is satisfied when the latter tends to zero at a polynomial rate, i.e., $\varphi(t)= O(t^{-\tau}), $ for all $\tau > 2$, as in the case of mixing random variables.\\
	Assumption A6 requires that the instruments  and  explanatory variables be bounded uniformly on $i$ and $n$. In addition, when the instruments depend on $\theta$ and $g(\cdot)$, they are also uniformly bounded  with respect to these parameters. The compactness condition in Assumption A7 is standard, and the convexity is somewhat unusual; however, it is reasonable in most applications. Condition A8 is necessary  to ensure the identification of the true parameters $\theta_0$. Assumption A9 requires  the standard deviations of the errors to be uniformly bounded away from zero with bounded derivatives. This  has been considered by \cite{pinkse1998contracting}. Assumption A10  is classic (\cite{pinkse1998contracting}) and required in the proof of Theorem~\ref{th2}. Those  authors noted that in their model (without a non-parametric component), when the autoregressive parameter $\lambda_0 = 0$,  $B_2(\theta_0)$ is not invertible, regardless of the choice of $M_n$. This is also the case in our context
	because for each $g_\theta(z)$  solution of  (\ref{partialH}), $\theta\in\Theta$ and $z\in \mathcal{Z}$, we have
	\begin{equation*}
	\frac{\partial g_{\theta}}{\partial \beta }(z)=-\frac{E\left( \left.\Gamma_{jn}(\theta, g_{\theta}(z)) X_{jn}\right\vert Z_{jn}=z \right)}{E\left( \left.\Gamma_{jn}(\theta, g_{\theta}(z)) \right\vert Z_{jn}=z \right)},
	\end{equation*}
	and 
	\begin{eqnarray*}
		\frac{\partial g_{\theta}}{\partial \lambda }(z)&=&\frac{v_{jn}^{'}(\lambda)}{v_{jn}(\lambda)}\frac{E\left( \left.\Gamma_{jn}(\theta, g_{\theta}(z))\left(X_{jn}^{T}\beta+g_{\theta}(z)\right)\right\vert Z_{jn}=z \right)}{E\left( \left.\Gamma_{jn}(\theta, g_{\theta}(z)) \right\vert Z_{jn}=z \right)}\\
		&=& \frac{v_{jn}^{'}(\lambda)}{v_{jn}(\lambda)} \left(g_{\theta}(z)-\beta^{T}	\frac{\partial g_{\theta}}{\partial \beta }(z)\right),
	\end{eqnarray*}
	where $v_{jn}^{'}(\lambda)=\frac{d }{d\lambda}v_{jn}(\lambda)=v_{jn}(\lambda)\left[W_nS_n^{-1}(\lambda)V_n(\lambda)\right]_{jj}$,  $$\Gamma_{jn}(\cdot)=\Lambda^{'}(G_{jn}(\cdot))\left[Y_{jn}-\Phi(G_{jn}(\cdot))\right]-\Lambda\left(G_{jn}(\cdot)\right)\phi\left(G_{jn}(\cdot)\right)$$
	and $\Lambda(\cdot)=\phi(\cdot)/(1-\Phi(\cdot))\Phi(\cdot)$. However
	$$\left.\frac{\partial g_{\theta}}{\partial \lambda }(z)\right\vert_{\lambda=0}=0 \qquad\mbox{because}\qquad v_{jn}^{'}(0)=0,$$
	then $B_2(\theta_0)$ will be singular when $\lambda_0=0$.
	
\end{remark}
With these assumptions in place, we are able to give some asymptotic results. The weak consistencies of the proposed estimators are given in the following two results.
The first theorem and corollary below establish the consistency of our estimators, whereas the second theorem addresses the question of convergence to a normal distribution of the parametric component when it is properly standardized.

\begin{theorem}\label{th1}
	Under Assumptions A1-A10, we have 
	\begin{equation*}
	\hat{\theta}-\theta_{0} =o_{p}(1).
	\end{equation*}
\end{theorem}
\begin{corollary}\label{th3}
	If the assumptions of  Theorem~ \ref{th1} are satisfied, then we have 
	\begin{equation*}
	\left\Vert \hat{g}_{\hat{\theta}}-g_{0}\right\Vert =o_{p}(1).
	\end{equation*}
\end{corollary}
\textbf{Proof of Corollary~\ref{th3}}
Note that
\begin{eqnarray*}
	\left\Vert \hat{g}_{\hat{\theta}}-g_{0}\right\Vert &\leq& \Vert \hat{g}_{\hat{\theta}}-g_{\hat{\theta}}\Vert+\Vert g_{\hat{\theta}}-g_{0}\Vert \\
	&\leq& \sup_{\theta}\Vert \hat{g}_{\theta}- g_{\theta}\Vert +\sup_{\theta}\left\Vert \frac{\partial g_{\theta}}{\partial \theta} \right \Vert \Vert \hat{\theta} - \theta_0 \Vert = o_p(1),  
\end{eqnarray*}
since, by the assumptions of Theorem~\ref{th1}, $\sup_{\theta}\Vert \hat{g}_{\theta}- g_{\theta}\Vert=o_p(1)$ and $\sup_{\theta}\left\Vert \frac{\partial g_{\theta}}{\partial \theta} \right \Vert < \infty$.\\
The following gives an asymptotic normality result of $\hat{\theta}$. 
\begin{theorem}\label{th2}
	Under assumptions A1-A10, we have 
	\begin{equation*}
	\sqrt{n}\left(\hat{\theta}-\theta_{0} \right)\rightarrow \mathcal{N}\left(0, \Omega(\theta_0) \right)
	\end{equation*}
\end{theorem}

\begin{remark}
	In practice, the previous asymptotic normality result can be used to construct asymptotic confidence intervals and build hypothesis tests when a consistent estimate of the asymptotic covariance matrix $\Omega(\theta_0)$ is available. To estimate this matrix, let us follow the idea of \cite{pinkse1998contracting} and define the estimator $$\Omega_n(\hat{\theta})=\left\{ B_{2n}(\hat{\theta})\right\}^{-1}\left\{\left.\frac{d  }{d \theta}S^{T}_n\left(\theta,\hat{g}_{\theta}\right)\right\vert_{\theta=\hat{\theta}}\right \}M_n B_{1n}(\hat{\theta})M_n \left\{\left.\frac{d }{d \theta}S_n\left(\theta,\hat{g}_{\theta}\right)\right\vert_{\theta=\hat{\theta}}\right\} \left\{ B_{2n}(\hat{\theta})\right\}^{-1},$$
	with 
	\begin{equation*}
	B_{1n}(\theta)=nS_n(\theta,\hat{g}_{\theta})S_n^{T}(\theta,\hat{g}_{\theta})\qquad \mathrm{and}\qquad 	B_{2n}(\theta)=\left\{\frac{d }{d \theta}S_n^{T}\left(\theta,\hat{g}_{\theta}\right)\right \}M_n\left\{\frac{d }{d \theta}S_n\left(\theta,\hat{g}_{\theta}\right)\right\}. 
	\end{equation*}
	The consistency of $\Omega_n(\hat{\theta})$ will be based on that  of $B_{1n}(\hat{\theta})$ and $B_{2n}(\hat{\theta})$, the estimators of $B_{1}(\theta_0)$ and $B_{2}(\theta_0)$, respectively. 
	Note that the  consistency of $B_{2n}(\hat{\theta})$ is relatively easy to establish. On the other hand, that of  $B_{1n}(\hat{\theta})$ asks for additional assumptions and an adaption of the proof of Theorem~3 of \cite[][p.134]{pinkse1998contracting} to our case; this is of interest to future research.  
\end{remark}

\section{Computation of the estimates}\label{sec4}
The aim of this section is to outline in detail how the regression parameters $\beta$, the spatial auto-correlation parameter $\lambda$ and the non-linear function $g_{\theta}$ can be estimated. 
We begin with the computation of $\hat{g}_{\theta}(z)$, which will play a crucial role in what follows.
\subsection{Computation of the estimate of the non-parametric component }
An iterative method is needed to compute the $\hat{g}_{\theta}(z)$ solution of (\ref{eq1}) for each fixed $\theta\in \Theta$ and $z\in \mathcal{Z}$. For fixed $\theta^{T}=(\beta,\lambda) \in \Theta$ and $z\in \mathcal{Z}$, let $\eta_\theta=g_{\theta}(z)$ and $\psi(\eta;\theta,z)$ denote the left-hand side of (\ref{eq1}), which can be rewritten as 
\begin{equation}
\psi(\eta;\theta,z)
=\sum_{i=1}^{n}\left[v_{in}(\lambda)\right]^{-1}\Lambda\left(G_{in}(\theta,\eta)\right)\left[Y_{in}-\Phi(G_{in}(\theta,\eta))\right]K\left(\frac{z-Z_{in}}{b_n}\right).
\end{equation}
Consider the Fisher information:
\begin{eqnarray}
\varPsi(\eta_\theta; \theta,z)&=& E_0\left(\left.\left.\frac{\partial}{\partial\eta}\psi(\eta; \theta,z)\right\vert_{\eta=\eta_\theta}\right\vert \left\{(X_{in},Z_{in}),\, 1\leq\ i\leq n,\, n=1,\ldots\right\}\right) \nonumber \\
&=&-\sum_{i=1}^{n}\left[v_{in}(\lambda)\right]^{-2}\Lambda\left(G_{in}(\theta,\eta_\theta)\right)\phi\left(G_{in}(\theta,\eta_\theta)\right)K\left(\frac{z-Z_{in}}{b_n}\right)	+ \nonumber \\
&&
\quad+\sum_{i=1}^{n}\left[v_{in}(\lambda)\right]^{-2}\Lambda^{'}\left(G_{in}(\theta,\eta_\theta)\right)\left[\Phi\left(G_{in}(\theta_0,\eta_0)\right)-\Phi\left(G_{in}(\theta,\eta_\theta)\right)\right]K\left(\frac{z-Z_{in}}{b_n}\right).
\label{FI}	
\end{eqnarray}
Note that the second term in the RHS of (\ref{FI}) is negligible when $\theta$ is near the true parameter $\theta_0$.  \\
Because $\psi(\eta;\theta,z)=0$ for $\eta=\hat{g}_{\theta}(z)$, an initial estimate $\tilde{\eta}$ can be updated to $\eta^\dagger$ using Fisher's scoring method: 
\begin{equation}
\eta^\dagger=\tilde{\eta}-\frac{\psi(\tilde{\eta};\theta,z)}{\varPsi(\tilde{\eta}; \theta,z)}.
\label{SFI}
\end{equation} 
The iteration procedure (\ref{SFI}) requests some starting value $\tilde{\eta}=\tilde{\eta}_0$ to ensure convergence of the algorithm. To this end, let us adapt  the approach of \cite{severini1994quasi}, which consists of supposing that for fixed $\theta\in \Theta$, there exists a $\tilde{\eta}_0$ satisfying 
$G_{in}(\theta,\tilde{\eta}_0)=\Phi^{-1}(Y_{in})$ for $i=1,\ldots,n$. Knowing that 
$G_{in}(\theta,\tilde{\eta}_0)=\left( v_{in}(\lambda)\right)^{-1}\left(X_{ni}^{T}\beta+ \tilde{\eta}_0\right)$, we have   
$\tilde{\eta}_0=v_{in}(\lambda)\Phi^{-1}(Y_{in})-X_{in}^{T}\beta$. Then, (\ref{SFI}) can be updated using the following initial value:
\begin{equation*}
\eta^{\dagger}_0=\tilde{\eta}_0-\frac{\psi(\tilde{\eta_0};\theta,z)}{\varPsi(\tilde{\eta_0}; \theta,z)}=\frac{\sum_{i=1}^{n}\left[v_{in}(\lambda)\right]^{-1}\Lambda(C_{in})\phi(C_{in})\left[C_{in}-\left[v_{in}(\lambda)\right]^{-1} X^T_{in}\beta\right]K\left(\frac{z-Z_{in}}{b_n}\right)}{\sum_{ i=1}^{n}\left[v_{in}(\lambda)\right]^{-2}\Lambda(C_{in})\phi(C_{in})K\left(\frac{z-Z_{in}}{b_n}\right)},
\end{equation*}
where  $C_{in}=\Phi^{-1}(Y_{in})$, $i=1,\ldots,n$, is computed using a slight adjustment because  $Y_{in}\in \{0,1\}$. \\
With this initial value, the algorithm iterates until convergence.
\subsubsection*{Selection of the bandwidth}

A critical step (in non- or semi-parametric models) is the choice of the bandwidth parameter $b_n$, which is usually selected by applying some cross-validation approach. The latter was adapted by \cite{su2012semiparametric} in the case of a spatial semi-parametric model. Because cross-validation may be very time consuming, which is true in  the case of our model, we adapt the following approach used in \cite{severini1994quasi} to achieve greater flexibility:
\begin{itemize}
	\item[1.]  Consider the linear regression of $C_{in}$ on $X_{in},\; i=1,\ldots,n$, without an intercept term, and let $R_{1n},\ldots, R_{nn}$ denote the corresponding  residuals.
	\item[2.] Since we expect $\mathbb{E}(R_{in}\vert Z_{in}=z)$ to have similar smoothness properties as $g_0(.)$, the optimal bandwidth $b_n$  is that of the non-parametric regression of the $\{R_{in}\}_{i=1,\cdots,n}$ on $\{Z_{in}\}_{i=1,\cdots,n}$, chosen by applying any non-parametric regression bandwidth selection method. For that, we use the cross-validation method in the \textit{np} R Package. 
\end{itemize}

\subsection{Computation of $\hat{\theta}$}
The parametric component $\beta$ and the spatial autoregressive parameter $\lambda$ are computed as mentioned above
by a GMM approach based on some instrumental variables $\xi_{n}$ and the weight matrix $M_n$.
The choices of these instrumental variables  and  weight matrix $M_n$ are as follows.\\
Because $\psi(\hat{g}_{\theta}(z);\theta,z)=0$, if we differentiate the latter with respect to $\beta$ and $\lambda$, we have  
\begin{equation*}
\frac{\partial }{\partial \beta}\hat{g}_{\theta}(z)=-\frac{\sum_{i=1}^{n}\left[v_{in}(\lambda)\right]^{-2} \Delta_{in}(\theta,z)X_{in}K\left(\frac{z-Z_{in}}{b_n}\right)}{\sum_{i=1}^{n}\left[v_{in}(\lambda)\right]^{-2}\Delta_{in}(\theta,z)K\left(\frac{z-Z_{in}}{b_n}\right)},
\end{equation*}
and 
\begin{eqnarray*}
	\frac{\partial }{\partial \lambda}\hat{g}_{\theta}(z)&=&\frac{\sum_{i=1}^{n}\left[v_{in}(\lambda)\right]^{-1}v_{in}^{'}(\lambda) \Delta_{in}(\theta,z)\left[X_{in}^{T}\beta+\hat{g}_{\theta}(z)\right]K\left(\frac{z-Z_{in}}{b_n}\right)}{\sum_{i=1}^{n}\left[v_{in}(\lambda)\right]^{-2}\Delta_{in}(\theta,z)K\left(\frac{z-Z_{in}}{b_n}\right)}\\
	&&\qquad + \frac{\sum_{i=1}^{n}\left[v_{in}(\lambda)\right]^{-2}v_{in}^{'}(\lambda) \Lambda\left(G_{in}(\theta,\hat{g}_{\theta}(z))\right)\left[Y_{in}-\Phi\left(G_{in}(\theta,\hat{g}_{\theta}(z))\right)\right]K\left(\frac{z-Z_{in}}{b_n}\right)}{\sum_{i=1}^{n}\left[v_{in}(\lambda)\right]^{-2}\Delta_{in}(\theta,z)K\left(\frac{z-Z_{in}}{b_n}\right)},
\end{eqnarray*}
with 
\begin{equation*}
\Delta_{in}(\theta,z)=\Lambda^{'}\left(G_{in}(\theta,\hat{g}_{\theta}(z))\right)\left[Y_{in}-\Phi\left(G_{in}(\theta,\hat{g}_{\theta}(z))\right)\right]-\Lambda\left(G_{ni}(\theta,\hat{g}_{\theta}(z))\right)\phi\left(G_{in}(\theta,\hat{g}_{\theta}(z))\right).
\end{equation*} 
Then, the previous result is used to define the following instrumental variables:
\begin{equation*}
\xi_{in}(\theta,\hat{g}_{\theta})=\frac{\partial G_{in}(\theta,\hat{\eta}_i)}{\partial\theta}+\frac{\partial G_{in}(\theta,\hat{\eta}_i)}{\partial\eta}\, \frac{\partial }{\partial \theta}\hat{g}_{\theta}(Z_{in}),
\end{equation*}
with $\hat{\eta}_i=\hat{g}_{\theta}(Z_{in})$.\\
For the weight matrix, we use (as in \cite{pinkse1998contracting})
$M_n=I_{q}$ with  $q=p+1$. Then,  the obtained GMM estimator of $\theta$ with this choice of $M_n$ is equal to the pseudo-profile maximum likelihood estimator of  $\theta$, accounting  only for the spatial heteroscedasticity.\\ The final step is to plug in the GMM estimator $\hat{\theta}$ to obtain $\hat{g}_{\hat{\theta}}$.

\section{Simulation study}\label{sec5}
In this section, we study the performance of the proposed model based on some numerical results, which highlight the importance of considering the spatial dependence and the partial linearity. We simulated some semi-parametric models and estimated them using our proposed method, i.e., the method that does not account for the spatial dependence (using the same estimation procedure above based on the partially linear probit  model (PLPM)), and using a fully linear SAE probit (LSAEP) method. The latter method can account for the  spatial dependence but ignores the partial linearity. The \textit{ProbitSpatial} R package \citep[][]{Martinetti} is used to provide estimates for the LSAEP model.  We generate observations from the following spatial latent partial linear model: 
\begin{eqnarray}
Y_{in}^{*}&=&\beta_1X_{in}^{(1)}+\beta_2X^{(2)}_{in}+g(Z_{in})+U_{in};\qquad Y_{in}=\mathbb{I}(Y_{in}^{*}>0),\, i=1,\ldots,n\\
U_{n}&=&(I_n-\lambda W_n)^{-1}\varepsilon_n
\end{eqnarray} 
where $U_{n}\sim \mathcal{N}(0, I_n)$ and  $W_n$ is the spatial weight matrix associated with $n$ locations chosen randomly in a $60\times 60$  regular grid based on the $6$ nearest neighbours of each unit. 
To observe the effect of partial linearity when we compare our estimation procedure to that based on LSAEP models, we will consider the following two cases:
\begin{description}
	\item[Case 1:] The explanatory variables $X^{(1)}$ and $X^{(2)}$ are generated as pseudo $\mathcal{B}(0.7)$ and  $\mathcal{U}[-2, 2]$, respectively, and the other explanatory variable $Z$ is equal to the sum of $48$ independent random variables, each uniformly distributed over $[-0.25, 0.25]$. Here, we use the non-linear function  $g(t)=t+2\cos(0.5\pi t)$. 
	\item[Case 2:]  The explanatory variables $X^{(1)}$, $X^{(2)}$ and $Z$ are  generated as pseudo  $\mathcal{N}(0, 1)$, and  we considerer the linear function  $g(t)=1+0.5t$.
\end{description}
We take $\beta_1=-1$, $\beta_2=1$ and different values of the spatial parameter $\lambda$, that is, $\lambda\in \{0.2, 0.5, 0.8\}$. The bandwidth $b_n$ is selected using \cite{severini1994quasi}'s approach  detailed previously with $C_{ni}=\Phi^{-1}\left(0.9Y_{ni}+0.1(1-Y_{ni})\right), i=1,\ldots,n$. A Gaussian kernel will be considered: $K(t)=(2\pi^{-1/2})\exp(-t^2/2)$. As mentioned above, the instrumental variables are the trivial choice, and the weight matrix $M_n=I_{3}$ is the identity matrix.

The two studied cases are replicated $200$ times for a sample size $n=200$, and the results are presented in Tables \ref{Tab4.1} and \ref{Tab4.2}. In each table, the columns titles Mean, Median  and SD give the average,  median and standard deviation, respectively, over these $200$ replications associated with each estimation method. 

\begin{figure}
	\begin{tabular}{c c  c}
		$\lambda=0.2$  & $\lambda=0.5$&  $\lambda=0.8$\vspace{-0.1cm}\\
		\includegraphics[width=0.3\linewidth]{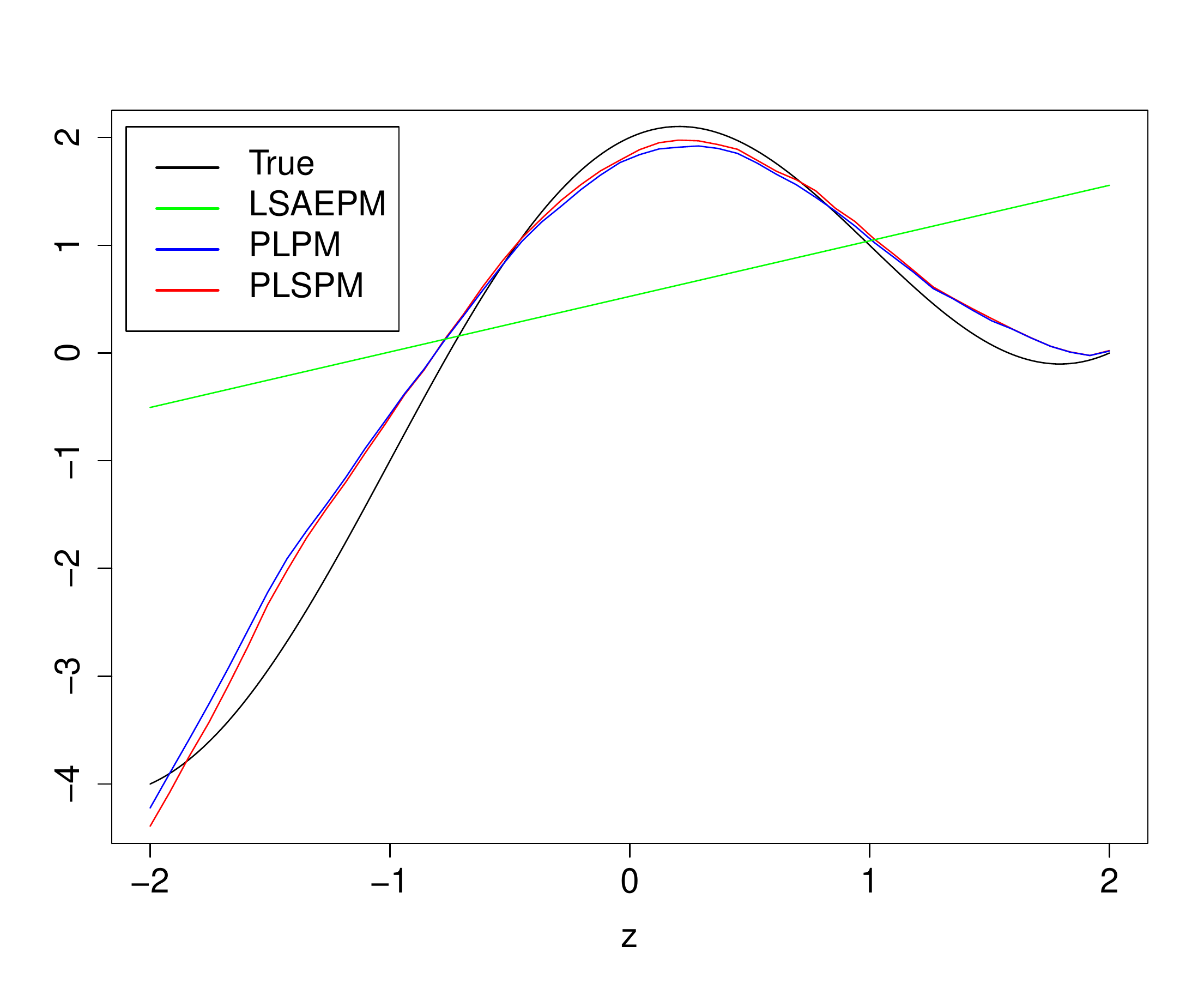} & \includegraphics[width=0.329\linewidth]{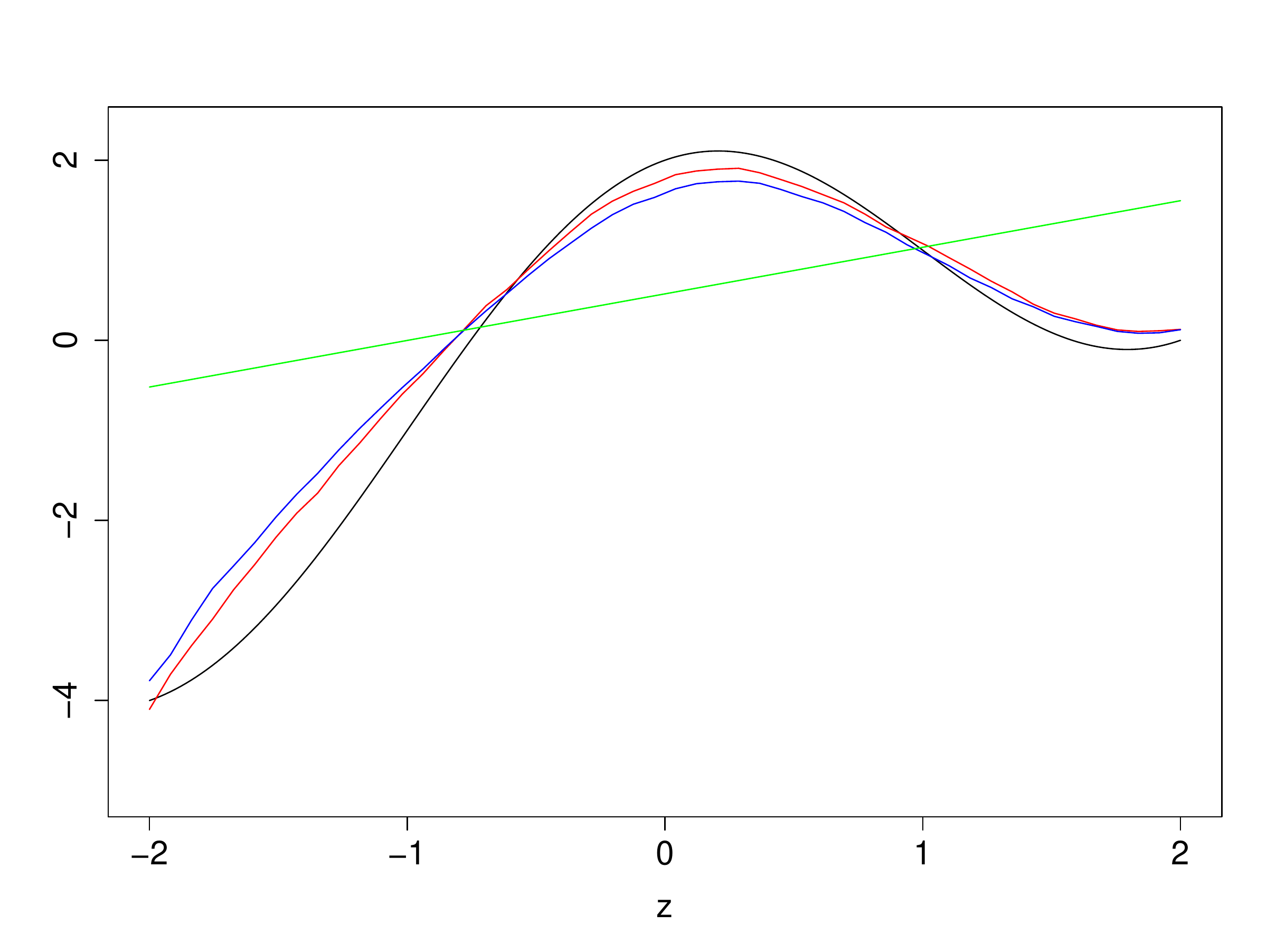} &  \includegraphics[width=0.329\linewidth]{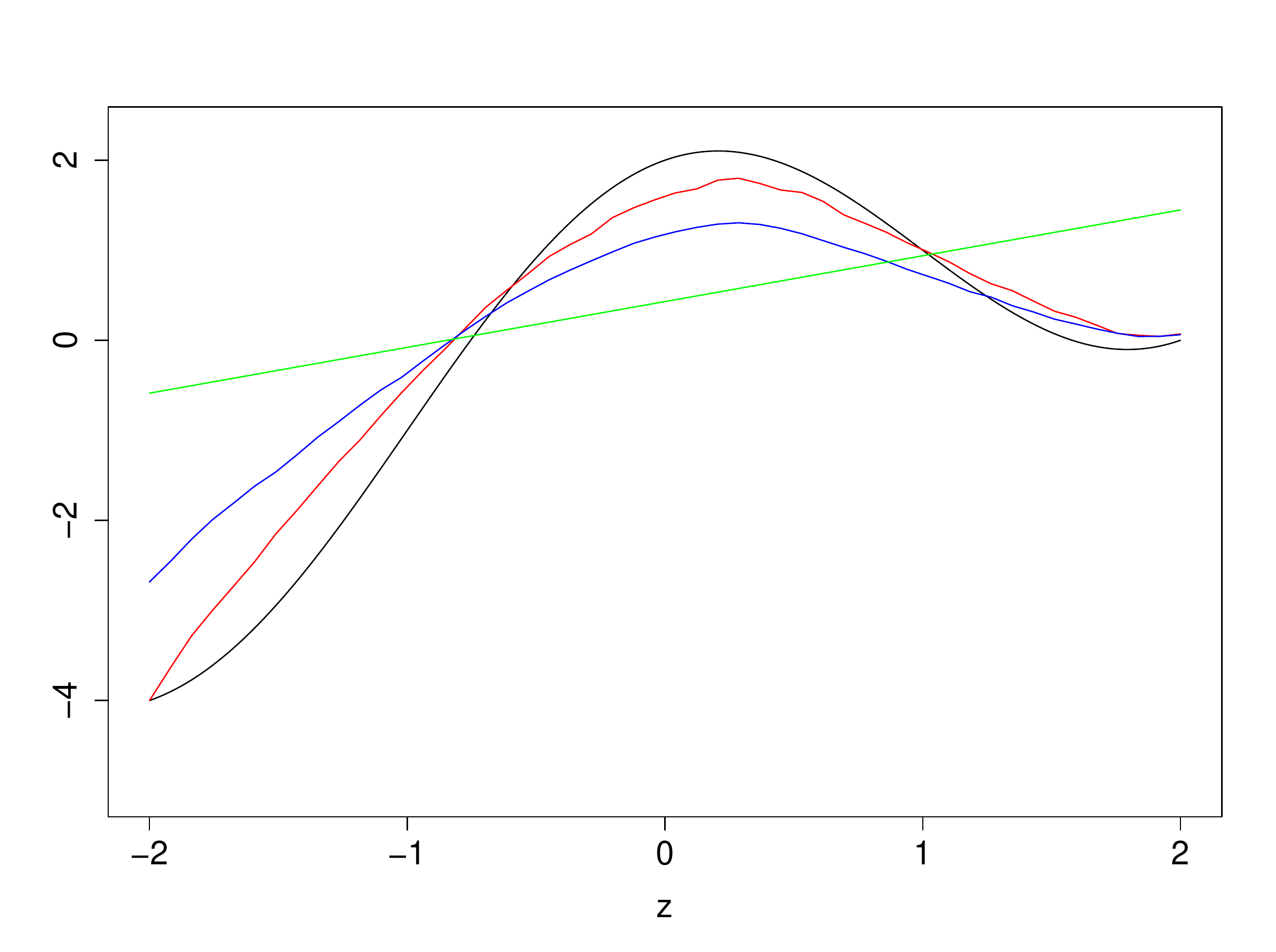}
	\end{tabular}
	\caption{Case 1 with $n=200$ and $200$ replications.}
	\label{Fig4.1}
\end{figure}

First, when we compare the estimators based on our approach  (PLSPM) with those  based on the LSAEP model, we notice that the latter yields more biased estimators of the coefficients  $\beta_1$ and $\beta_2$, in particular in Case 1.  It makes sense that ignoring the partial linearity (see also Figure~\ref{Fig4.1}) weakens the quality of the estimation of the coefficients $\beta_1$ and $\beta_2$. In Case 2, these two approaches yield similar results in term of consistency,  but our approach seems to be  less efficient.    

Second, note that for the two cases (Table~\ref{Tab4.1} and Table~\ref{Tab4.2}), the  LSAEP and PLPM estimates are similar in the case of low spatial dependence ($\lambda=0.2$). However, this is not the case for the large spatial dependence ($\lambda=0.8$) framework, where in this case the estimation procedure based on PLPM models yields inconsistent estimates of the parameters $\beta_1$ and $\beta_2$ and the smooth function $g(\cdot)$ (see the right panel in Figure~\ref{Fig4.1}). It makes sense that considering the spatial dependence does not allow one to find consistent estimates of the coefficients $\beta_1$ and $\beta_2$ and the smooth function $g(\cdot)$.\\ 
Note that this approach is less efficient; this can be realized when observing the differences between the mean and median (or the high values of the standard deviation) associated with our estimators in Tables~\ref{Tab4.1}-\ref{Tab4.2}. However, this is eventually due to the use  of the GMM approach with the trivial choice of the weight matrix $M_n=I_n$. In addition, when estimating the spatial parameter $\lambda$, our procedure yields biased estimators; this may be related to the considered  choice of IVs. Better choices of the weight matrix and instrumental variables have to be investigated in future research.

\begin{table}
	\centering
	\begin{tabular}{c l ccc c ccc c ccc  }
		\hline	
		\multirow{2}{*}{$\lambda $} & \multirow{2}{*}{Methods}&  \multicolumn{3}{l}{$\beta_1=-1$ }&  &\multicolumn{3}{l}{$\beta_2=1$ }&  &\multicolumn{3}{l}{$\lambda$ }	\\
		\cline{3-5}                             \cline{7-9}                        \cline{11-13}
		&                         & Mean   & Median & SD               &   & Mean & Median & SD             &  & Mean & Median & SD \\ \hline 
		\multirow{3}{*}{ 0.20}      & PLSPM                   & -1.08  & -1.00  & 0.53             &   & 1.07 & 0.99   & 0.33           &  & 0.09 & 0.00   & 0.29  \\
		& LSAEP                    & -0.67  & -0.69  & 0.25             &   & 0.67 & 0.66   & 0.11			&  & -0.04&  0.02  & 0.36  \\
		& PLPM                    & -0.98  & -0.99  & 0.32             &   &0.98 & 0.96    & 0.15           &  &       &       &         \\ \hline
		
		\multirow{3}{*}{ 0.50}      & PLSPM                   & -1.13  & -0.96  & 0.67             &   & 1.08 & 0.98   & 0.40           &  & 0.27 & 0.10   & 0.37  \\
		& LSAEP                   & -0.65  & -0.64  & 0.24             &   & 0.66 & 0.65   & 0.12			&  & 0.20 &  0.26  & 0.29  \\
		& PLPM                    & -0.90  & -0.88  & 0.30             &   &0.90  & 0.89   & 0.15           &  &       &       &         \\ \hline 
		
		\multirow{3}{*}{ 0.80}      & PLSPM                   & -1.12  & -0.86  & 0.86             &   & 1.08 & 0.89   & 0.55           &  & 0.53 & 0.71   & 0.39  \\
		& LSAEP                    & -0.57  & -0.56  &  0.25            &   & 0.61 & 0.60   & 0.12			&  & 0.60 &  0.61  & 0.10  \\
		& PLPM                    & -0.65  & -0.66  & 0.25             &   & 0.65 & 0.63   & 0.13           &  &       &       &         \\ \hline 
	\end{tabular} 
	\caption{Case 1 with $n=200$ and $200$ replications.}
	\label{Tab4.1}
\end{table} 
\begin{table}
	\centering
	\begin{tabular}{c l ccc c ccc c ccc  }
		\hline	
		\multirow{2}{*}{$\lambda $} & \multirow{2}{*}{Methods}&  \multicolumn{3}{l}{$\beta_1=-1$ }&  &\multicolumn{3}{l}{$\beta_2=1$ }&  &\multicolumn{3}{l}{$\lambda$ }	\\
		\cline{3-5}                             \cline{7-9}                        \cline{11-13}
		&                         & Mean   & Median & SD               &   & Mean & Median & SD             &  & Mean & Median & SD \\ \hline 
		\multirow{3}{*}{ 0.20}      & PLSPM                   & -1.12  & -1.05  & 0.32             &   & 1.13 & 1.06   & 0.30           &  & 0.26 & 0.05   & 0.31  \\
		& LSAEP                    & -1.08  & -1.06  & 0.19             &   & 1.09 & 1.07   & 0.20			&  & 0.02 & 0.17   & 0.47  \\
		& PLPM                    & -1.00  & -0.99  & 0.20             &   &0.99  & 0.98   & 0.14           &  &      &       &         \\ \hline
		
		\multirow{3}{*}{ 0.50}      & PLSPM                   & -1.08  & -1.03  & 0.37             &   & 1.06 & 0.99   & 0.31           &  & 0.30 & 0.18   & 0.31   \\
		&LSAEP                    & -1.06  & -1.06  & 0.21             &   & 1.05 & 1.01   & 0.19			&  & 0.40 &  0.48  & 0.29 \\
		& PLPM                    & -0.95  & -0.94  & 0.21             &   &0.93  & 0.91   & 0.18           &  &      &       &         \\ \hline 
		
		\multirow{3}{*}{ 0.80}      & PLSPM                   & -1.02  & -0.91  & 0.44             &   & 1.01 & 0.86   & 0.43           &  & 0.56 & 0.68   & 0.35  \\
		& LSAEP                    & -0.88  & -0.87  & 0.19             &   & 0.87 & 0.86   & 0.20			&  & 0.72 &  0.73  & 0.09  \\
		& PLPM                    & -0.66  & -0.65  & 0.15             &   & 0.66 & 0.65   & 0.16           &  &       &       &         \\ \hline 
	\end{tabular} 
	\caption{Case 2 with $n=200$ and $200$ replications.}
	\label{Tab4.2}
\end{table} 


\section*{Discussion}
In this manuscript, we have proposed a spatial semi-parametric probit model for identifying risk factors at onset and with spatial heterogeneity. 
The parameters involved in the models are estimated using weighted likelihood and generalized method of moment methods. A technique based on dependent random arrays  facilitates the estimation and derivation of asymptotic properties, which otherwise would have been difficult to perform due to the complexity introduced by the spatial dependence to the model and high-dimensional integration required by a full maximum likelihood approach.
Moreover, the technique yields consistent estimates through proper choices of the bandwidth, weight matrix, and instrumental variables.
The proposed models provide a general framework and tools for researchers and practitioners when addressing binary semi-parametric choice models in the presence of spatial correlation. Although they provide significant contributions to the body of knowledge, to the best of our knowledge, additional work  needs to be done.\\ 
As indicated, the weights are used to improve the efficiency and convergence. It would be interesting to develop criteria for the choices of optimal weights toward achieving  better performance. 
For instance, the performance may be improved by choosing, for instance, a weight matrix $M_n$ as a consistent estimator $B_{1n}(\hat{\theta})$ of the matrix 
$B_{1}({\theta_0})$. Another empirical choice could be the idea of continuously updating the GMM estimator (one-step GMM) used in \cite{pinkse2006dynamic}:
\begin{equation*}
M_n(\theta)=n^{-1}\sum_{i,j=1}^{n}\delta_{ij}\xi_{ni}\xi_{jn}^T\tilde{U}_{in}(\theta,\hat{g}_{\theta})\tilde{U}_{jn}(\theta,\hat{g}_{\theta})
\end{equation*} 
with  the weights 
\begin{equation*}
\delta_{ij}=\frac{\sum_{r=1}^{n}\tau_{ri}\tau_{rj}}{\left[\sum_{r=1}^{n}\tau_{ri}^2\sum_{r=1}^{n}\tau_{rj}^2\right]^{1/2}} \qquad\mbox{  for  } i,j=1,\ldots,n,
\end{equation*}  
where $\tau_{ij}$ is a number depending on $W_{nij}$. The nearer  $i$ is to $j$, the larger  $\tau_{ij}$ is.\\
Another topic of future research is in allowing some spatial dependency in the covariates (SAR models) and the response (endogenous models) for greater generality.
These topics will be of interest in future research.

\section{Appendix}\label{secApp}
\begin{proposition}\label{prop1}
	Under Assumptions A1-A6, for $\theta\in \Theta$ and $z\in \mathcal{Z}$, the functions $g_\theta(z)$ and $\hat{g}_{\theta}(z)$, solutions of (\ref{partialH}) and (\ref{eq1}), respectively, satisfy  
	\begin{enumerate}
		\item for all $i,j=0, 1, 2, \; i+j\leq 2$, 
		$$\frac{\partial^{i+j}}{\partial \theta_l^{i}\partial\theta_r^{j}}g_{\theta}(z)\qquad \mathrm{and } \qquad \frac{\partial^{i+j}}{\partial \theta_l^{i}\partial\theta_r^{j}}\hat{g}_{\theta}(z) \qquad \mbox{exist and are finite for all } 1\leq l,r\leq p+1.$$
		\item $\displaystyle \sup_{\theta\in \Theta}\left \Vert \hat{g}_{\theta}-g_{\theta}\right\Vert $, $\displaystyle\sup_{\theta\in \Theta}\max_{j=1,\ldots,p+1}\left \Vert  \frac{\partial}{\partial\theta_j}\left(\hat{g}_{\theta}- g_{\theta}\right)\right\Vert $ and   $\displaystyle\sup_{\theta\in \Theta}\max_{1\leq i ,j\leq p+1}\left \Vert \frac{\partial^{2}}{\partial\theta_i\partial \theta_j}\left(\hat{g}_{\theta}- g_{\theta}\right) \right\Vert$, 
	\end{enumerate}
	are all order $o_p(1)$ as $n\to \infty$.
\end{proposition}
Without loss of generality, the proof of this proposition is ensured by Lemma~\ref{l4}  in the  univariate case  i.e., $\Theta,\, \mathcal{Z} \subset \mathbb{R}$. \\

The following lemma is useful in the proof of Lemma~\ref{l4}. It is an extension of  Lemma~8 in \cite{severini1992profile} to spatially dependent data.

\begin{lemma}\label{l8}
	Let $\zeta_\theta (Y_i)$  denote a scalar function of $Y_{in}$, $i=1,\ldots,n,\; n=1,2,\ldots$, depending on a scalar parameter $\theta\in \Theta$, and for  $j=0,1,2$, let
	$$\zeta^{(j)}_{\theta}(Y_{in})=\frac{\partial^j}{\partial \theta^j} \zeta_\theta (Y_{in}), \qquad i=1,\ldots,n,\; n=1,2,\ldots$$
	Let $f_i(\cdot)$ denote the density of $Z_{in}$ (given in Assumption A2), and let $\bar{f}(z)=\frac{1}{n}\sum_{i=1}^{n}f_{i}(z)$. \\ 
	Assume that
	\begin{itemize}
		\item[ \textbf{H.1}] $\displaystyle \sup _{\theta}\sup _{1\leq i\leq n, n}  \left \vert\zeta_\theta ^{(j)}(Y_{in}) \right \vert < \infty$ for $ j=0,\ldots,3$.
		\item[\textbf{H.2}] For all $\theta\in \Theta$, $j=0,1,2$, and $1\leq i,l\leq n$: 
		\begin{equation}
		\left\vert \mathrm{Cov}\left(K_{in}(z),K_{ln}(z)\right)\right\vert\leq\left\{\mathrm{Var}(K_{in}(z))\mathrm{Var}(K_{in}(z))\right\}^{1/2}\varphi\left(\Vert s_i-s_l\Vert\right),
		\label{dep1}
		\end{equation}
		\begin{align}	
		&\left\vert \mathrm{Cov}\left(\zeta^{(j)}_{\theta}(Y_{in})K_{in}(z),\zeta^{(j)}_{\theta}(Y_{ln})K_{ln}(z)\right)\right\vert\leq\nonumber \\&\qquad\qquad\left\{\mathrm{Var}\left(\zeta^{(j)}_{\theta}(Y_{in})K_{in}(z)\right)\mathrm{Var}\left(\zeta^{(j)}_{\theta}(Y_{ln})K_{ln}(z)\right)\right\}^{1/2}\varphi\left(\Vert s_i-s_l\Vert\right),
		\label{dep2}
		\end{align}
		with $K_{in}(z)=K\left( (z-Z_{in})/b\right)$.		
	\end{itemize}
	Let $m_{\theta}(z)=\mathbb{E}\left(\zeta_{\theta} (Y_{in}) | Z_{in}=z \right)$ for $z\in \mathcal{Z}$, and assume that $\displaystyle \frac{\partial^j}{\partial \theta^j}m_\theta(\cdot)$ is continuous on $\mathcal{Z}$,  $j=0,1,2$.  \\ 
	
	For each fixed $\theta \in \Theta$ and $z\in \mathcal{Z}$, let the kernel estimator $\widehat{m}_{\theta}(z)$ of $m_{\theta}(z)$  be defined by 
	$$\widehat{m}_{\theta}(z)=\frac{\sum_{i=1}^{n}\zeta_{\theta} (Y_{in}) K_{in}(z)}{\sum_{i=1}^{n} K_{in}(z)}.$$
	If Assumptions A2, A4, and A5  are satisfied,  then
	$$\sup_{\theta\in \Theta}\sup_{z\in\mathcal{Z} } \left \vert \frac{\partial^{j} }{\partial \theta^j}  \widehat{m}_{\theta}(z)-  \frac{\partial^{j} }{ \partial \theta^j}  m_{\theta}(z) \right \vert =o_p(1),$$
	for $j=0,1,2$.
\end{lemma} 

Lemma \ref{l8} generalizes  Lemma 8 in \cite{severini1992profile} to spatially dependent data.

\subsection*{Proof of Lemma \ref{l8}}
We give the proof in the case where $j=0$, corresponding to the study of the uniform consistency of the kernel estimator of the regression function of $\zeta_{\theta}(Y_{in})$ on $Z_{in}$. The other cases are similar to this case and thus are omitted. \\
Let 
$$\widehat{v}_\theta (z)=\frac{1}{nb^d}\sum_{i=1}^{n}\zeta_{\theta} (Y_{in})K_{in}(z);\qquad \widehat{f}(z)=\frac{1}{nb^d}\sum_{i=1}^{n} K_{in}(z), $$
$$v_\theta (z)= m_{\theta}(z)\bar{f}(z).$$
We have to show that 
\begin{eqnarray}
\sup_{\theta}\sup_{z}\left \vert \widehat{v}_{\theta}(z) -v_{\theta}(z)\right\vert=o_p(1)
\label{convunif1}
\end{eqnarray}
and 
\begin{eqnarray}
\sup_{z}\left \vert \widehat{f}(z) -\bar{f}(z)\right\vert=o_p(1)
\label{convunif2}
\end{eqnarray}
We give the proof of $(\ref{convunif1})$, and that of $(\ref{convunif2})$ is similar. \\
\\
\subsection*{\underline{\textbf{Asymptotic behavior of} $\left \vert \widehat{v}_{\theta}(z) -v_{\theta}(z)\right\vert$}}
\medskip
Let us first consider the bias $\left \vert \mathbb{E}(\widehat{v}_{\theta}(z)) -v_{\theta}(z)\right\vert$. We have
\begin{eqnarray*}
	\mathbb{E}(\widehat{v}_{\theta}(z))&=& (nb^d)^{-1}\sum_{i=1}^{n}\int  K\left( \frac{z-u}{b}\right)m_{\theta}(u)f_i(u)du \\
	&=& b^{-d}\int v_{\theta}(u)K\left( \frac{z-u}{b}\right)du;\\
	&=&\int v_{\theta}(z-bu)K(u)du
\end{eqnarray*}
thus,
\begin{equation*}
\mathbb{E}(\widehat{v}_{\theta}(z))-v_{\theta}(z)= \int \left( v_{\theta}(z-bu)-v_{\theta}(z)\right)K(u)du=o(1)
\end{equation*}
by Assumption A4, the continuity of $f_i(\cdot)$ (see A2) and $m_\theta(\cdot)$, and the compactness of $\mathcal{Z}$.  
Clearly, the bias term does not depend on $\theta$ or $z$.\\
Let us now treat  $\left\vert \widehat{v}_{\theta}(z)-\mathbb{E}(\widehat{v}_{\theta}(z))\right\vert$. Consider  the sum of variances 
\begin{equation*}
\mathbf{S}_{n}= (nb^d)^{-2} \sum_{i=1}^{n}\mathrm{Var}\left( \zeta_{\theta} (Y_{in})K_{in}(z) \right).\end{equation*}
We have
\begin{eqnarray}
\mathrm{Var}\left( \zeta_{\theta} (Y_{in})K_{in}(z) \right)&\leq &  \mathbb{E}\left(\zeta_{\theta}^{2} (Y_{in}) K_{in}^{2}(z)\right)\nonumber\\
&\leq & C  \mathbb{E}\left( K_{in}^2(z)\right)= Cb^d \sum_{i=1}^{n} \int  K^{2}(u)f_{i}(z-ub) du \nonumber\\
&=& C b^d \sup_{u} \left \vert  K(u)\right \vert ^{2} \int f_i(z-ub) du= C b^d\, \sup_{u} \left \vert  K(u)\right \vert ^{2},
\label{VarTK}
\end{eqnarray}	
because $\zeta_{\theta}(Y_{in})$ is bounded uniformly on $i$ and $\theta$ by assumption {\bf H.1}, $\int f_i(z-ub)du\leq C$ (see assumption A2) and  $\sup_{u} \left \vert  K(u)\right \vert ^{2}< \infty$ (see Assumption A4 and the compactness of $\mathcal{Z}$). Then, we have
\begin{equation}
\mathbf{S}_{n}= O\left( (nb^d)^{-1}\right).
\label{Sn}
\end{equation}
Now, consider the  covariance term
\begin{equation*}
\mathbf{R}_{n}=\displaystyle (nb^d)^{-2}\sum_{i=1}^{n}\sum_{\underset{j\neq  i}{j=1}}^{n}\mbox{Cov}\left(\zeta_{\theta} (Y_{in})K_{in}(z),\zeta_{\theta} (Y_{jn})K_{jn}(z) \right).\end{equation*}  
Let us partition the spatial locations of the observations using
\begin{equation*}
D_{n}=\{1\leq i,j\leq n\, : \, \rho < \Vert s_i -s_j \Vert \leq c_n\}
\end{equation*}
with  $c_n$ being the sequence of integers going to $\infty$, and let $\bar{D}_{n}$ denote  the complement of $D_{n}$ in the set of locations $\{s_i,\, i=1,...,n\}$. 
\\ On the one hand, let 
\begin{equation*}
\mathbf{R}_n^{(1)}=(n\, b^d)^{-2}\,\sum_{i,j \in D_n} \left \vert \mathrm{Cov}\left(\zeta_{\theta} (Y_{in})K_{in}(z),\zeta_{\theta} (Y_{jn})K_{jn}(z) \right)\right \vert = (n\, b^d)^{-2}\,\sum_{i,j \in D_n} \vert A-B \vert,
\end{equation*} 
with 
\begin{eqnarray*}
	\vert A\vert &=&\left \vert \mathbb{E} \left(\zeta_{\theta} (Y_{in})K_{in}(z) \zeta_{\theta} (Y_{jn})K_{jn}(z) \right)\right \vert \\
	&\leq & C \left \vert \int  K\left(\frac{z-u}{b}\right)K\left(\frac{z-v}{b}\right)f_{i,j}(u,v)dudv\right\vert\\
	&\leq &C\, b^{2d} \left \vert\int K(u)K(v)f_{i,j}(z-bu,z-bv)dudv\right\vert\\ 
	&\leq & C b^{2d}\left(\sup_{u}\vert K(u)\vert\right)^2\left \vert\int f_{i,j}(z-bu,z-bv)dudv\right\vert=Cb^{2d},
\end{eqnarray*}
by Assumption {\bf H.1}, $\sup_{u}\vert K(u)\vert< \infty$ (Assumption A4 and the compactness of $\mathcal{Z}$), with $f_{i,j}$ being the joint density (Assumption A2 and the compactness of $\mathcal{Z}$).\\
Note that the second term $B$ is  
\begin{equation*}
B=\mathbb{E} \left(\zeta_{\theta} (Y_{in})K_{in}(z) \right) \mathbb{E} \left( \zeta_{\theta} (Y_{jn})K_{jn}(z) \right)
\end{equation*}  
\noindent  
Using similar arguments as above,  we have $\vert B\vert \leq Cb^{2d}$ by Assumptions  A2 and A4, the compactness of $\mathcal{Z}$ and the continuity of $m_{\theta}(\cdot)$.
Thus, we have
\begin{eqnarray}
\mathbf{R }_{n}^{(1)}& \leq & C n^{-2} \sum_{i,j \in D_n} \leq C \frac{c_n^{2}-\rho^{2}}{n}=O\left( \frac{c_n^2}{n}\right).
\label{R1}
\end{eqnarray}
On the other hand, let 

\begin{equation*}
{\bf R}_n^{(2)}= (n\, b^d)^{-2}\,\sum_{i,j \in \bar{D}_n} \left \vert \mathrm{Cov}\left(\zeta_{\theta} (Y_{in})K_{in}(z),\zeta_{\theta} (Y_{jn})K_{jn}(z) \right)\right \vert.
\end{equation*} 
By Assumption \textbf{H.2} combined with (\ref{VarTK}), we have  for all $\theta\in \Theta$ and $i,j=1,\ldots,n$,
\begin{equation*}
\left \vert \mathrm{Cov}\left(\zeta_{\theta} (Y_{in})K_{in}(z),\zeta_{\theta} (Y_{jn})K_{jn}(z) \right)\right \vert \leq C\,b^d\varphi(\Vert s_i - s_j\Vert).
\end{equation*}
Then, we have 
\begin{equation} 
\mathbf{R}_n^{(2)}\leq C(n\, b^{d})^{-1}  \sum_{i> c_n/\rho} i\varphi(i \rho).
\end{equation}
Thus, we derive the following result:
\begin{equation}
\mathbf{R}_{n}=\mathbf{R}_n^{(1)}+\mathbf{R}_n^{(2)} = O\left ( n^{-1}\left\{c_n^{2} + b^{-d}\sum_{i> c_n/\rho} i\varphi(i \rho)\right\} \right). 
\end{equation}

The following steps of the proof are inspired by the proof of Lemma~8 in  \cite{severini1992profile} (p. 1800--1801).
Let  
$$\tilde{v}_{\theta}(z)=\frac{1}{n}b^{-d} \sum_{i=1}^{n}\left \{ \zeta_{\theta}(Y_{in})K_{in}(z)-\mathbb{E}\left(\zeta_{\theta}(Y_{in})K_{in}(z)\right)\right\}.$$
For some $\epsilon >0$, Markov's inequality yields 
\begin{eqnarray}
\mathbb{P}\left(\left \vert \tilde{v}_{\theta}(z) \right \vert > \epsilon \right) &\leq & \frac{\mathbf{R}_{n}+\mathbf{S}_{n}}{\epsilon^{2}}.
\end{eqnarray}	

Now, let $\theta_1$ and $\theta_2$ be two elements in $\Theta$; because $\displaystyle\mathbb{E}\left(\sup_{\theta, 1\leq i\leq n, n} \vert \zeta_{\theta}^{(1)} (Y_{in}) \vert \right) < \infty $ (by \textbf{H.1}), there exists a random triangular array  \citep[see ][p.1801]{severini1992profile} $\left\{W^{(1)}_{in},\, 1\leq i\leq n, \, n=1,2\ldots \right\}$ not depending on $\theta_{1}$ and $\theta_2$ such that $\sup_{1\leq i\leq n, \,n}\mathbb{E}\left( \vert W_{in}^{(1)}\vert \right)< \infty $  and 
$$\sup_{z}\left \vert \tilde{v}_{\theta_1}(z)-\tilde{v}_{\theta_2}(z)\right \vert \leq \sup_{z} \left \vert K(z) \right \vert \frac{\vert \theta_2 - \theta_1 \vert }{b^d} \frac{1}{n} \sum_{i=1}^{n}W_{in}^{(1)}.$$
Similarly,  for all $z^{(1)} $ and $z^{(2)}$ in $\mathcal{Z}$, there exists a random triangular array \\ $\left\{W^{(2)}_{in},\, 1\leq i\leq n, \, n=1,2\ldots \right\}$ not depending on $z^{(1)}$ and $z^{(2)}$ such that $\sup_{1\leq i\leq n, \,n}\mathbb{E}\left( \vert W_{i}^{(2)}\vert \right)< \infty $ and 
$$\sup_{\theta}\left \vert \tilde{v}_{\theta}(z^{(2)})-\tilde{v}_{\theta}(z^{(1)})\right \vert \leq C \frac{\Vert z^{(2)} - z^{(1)} \Vert }{b^{d+1}} \frac{1}{n} \sum_{i=1}^{n}W_{in}^{(2)},$$
because  $K(\cdot)$  is Lipschitzian (see Assumption {\bf H.2}).\\
Hence, there exists a random triangular array $\left\{W_{in},\, 1\leq i\leq n, \, n=1,2\ldots \right\}$ such that $\sup_{1\leq i\leq n, \,n}\mathbb{E}\left( \vert W_{in}\vert \right)< \infty $ and 
\begin{eqnarray*}
	\sup_{\Vert z^{(2)} -z^{(1)}\Vert < \delta_1}\sup_{\vert \theta_2 - \theta_1 \vert < \delta_2}\left \vert \tilde{v}_{\theta_2}(z^{(2)})-\tilde{v}_{\theta_1}(z^{(1)})\right \vert &\leq& C  \left(b^{-d}\delta_2 +b^{-(d+1)}\delta_1 \right)\frac{1}{n} \sum_{i=1}^{n}W_{in},
\end{eqnarray*}
for some $\delta_1>0$, $\delta_2>0$ and  large  $n$.\\

Because $\mathcal{Z}$ is compact, one can define a real number $\delta_1>0$, an  integer $l_n$  such that $l_n\delta_1<C$  with $l_n=\lfloor \gamma_nb^{-(d+1)}\rfloor$ and  
$$\mathcal{Z}\subset \bigcup_{j=1}^{l_n} B(z^{(j)},\delta_1),$$
where $B(z,\delta)$ is the closed ball in $\mathbb{R}^{d}$ with center $z$ and radius $\delta>0$.\\ 
In addition, because $\Theta$ is  compact, one can cover it by $r_n=\lfloor\gamma_n b^{-d}\rfloor$ finite intervals of centers $\theta_{i}$ with the same half length $\delta_2=O(1/r_n)$.\\
With these coverings, we have 
\begin{align*}
& \mathbb{P}\left( \sup_{\theta,z}\left \vert \tilde{v}_{\theta}(z)\right \vert > \epsilon\right)\leq \mathbb{P}\left(\max_{j\leq r_n}\max_{k\leq l_n }\left \vert \tilde{v}_{\theta_j}(z^{(k)})\right \vert > \epsilon/2\right) \\
& \qquad\qquad\qquad \qquad \qquad \qquad+ \mathbb{P}\left(\sup_{\Vert z^{(2)} -z^{(1)}\Vert < \delta_1}\sup_{\vert \theta_2 - \theta_1 \vert < \delta_2}\left \vert \tilde{v}_{\theta_2}(z^{(2)})-\tilde{v}_{\theta_1}(z^{(1)})\right \vert > \epsilon/2 \right)\\
&  \qquad\qquad \qquad\qquad\quad\leq  \quad r_n\, l_n\,\mathbb{P}\left(\left \vert \tilde{v}_{\theta}(z)\right \vert > \varepsilon/2\right)+  C b^{-d} \left(\delta_2 +\delta_1 b^{-1}\right)\\
&\qquad\qquad \qquad\qquad\quad= \;   C\, r_n\, l_n(\mathbf{S}_n +\mathbf{R}_n)+ C b^{-d} \left(\delta_2 +\delta_1 b^{-1}\right)\\
&\qquad\qquad \qquad\qquad\quad:= \; I^{(1)}+I^{(2)}+I^{(3)},
\end{align*} 
where 
$$I^{(1)}= O\left(\frac{\gamma_n^2}{nb^{2d+1}}\left(c_n^2+b^{-d}\sum_{i> c_n/\rho}i\varphi(i\rho)\right)\right); \qquad  I^{(2)}=O\left(\gamma_n^{-1} \right) \,;\qquad
I^{(3)}=O\left(\frac{\gamma_n^2}{nb^{3d+1}}\right).$$
If we take $c_n=o(b^{-d/2})$ and $\gamma_n^2=o(nb^{3d+1})$, then $I^{(1)}, I^{(2)}$ and $I^{(3)}$ are all of order $o(1)$ by  Assumption A5 and by the fact that $\varphi(t)\to 0$ as $t\to \infty$ by Assumption A3.	This yields the proof. $\quad \square$
\begin{lemma}\label{l4}  For each $\theta \in \Theta$ and $z \in  \mathcal{Z}$, let
	\begin{equation*}
	H(\eta;\theta, z)= \mathbb{E}_{0}\left ( h_{in}^{\theta,\, \eta}(Y_{in} | X_{in},Z_{in}  )| Z_{in}=z\right ), \; 1\leq i\leq n, \; n=1,2,\ldots
	\end{equation*} 
	where $\eta =g(z), \, g\in \mathcal{G}$ and  $ h_{in}^{\theta,\, \eta}(\cdot |\cdot, \cdot)$ is defined in Assumption A3.\\
	
	\noindent\underline{\textbf{Condition I}}: For fixed but arbitrary $\theta_1 \in \Theta $ and $\eta_1 \in \varPi$ with $\varPi = g_0(\mathcal{Z})$, let 
	$$\vartheta (\theta,\eta) =\int h_{in}^{\theta,\eta}(y|x\, ,z) \exp(h_{in}^{\theta_1,\eta_1}(y|x\, ,z)) dy , \qquad \theta \in \Theta, \; \eta \in \varPi,(x,z)\in \mathcal{Z}\times\mathcal{Z}$$ 
	where  $\{\exp (h_{in}^{\theta,\eta}(y|x\, ,z)), \theta \in \Theta, \, \eta \in \varPi\}$  denotes the family of conditional density functions (indexed by the parameters  $\theta$ and $\eta$) of  $Y_{in}$ given $(X_{in},Z_{in})=(x,z)\in \mathcal{X}\times\mathcal{Z}$.
	For each  $\theta \neq \theta_1$, assume that
	$$\vartheta(\theta, \eta) < \vartheta(\theta_1, \eta_1). $$  
	\underline{\textbf{Condition S}}: 
	Let $\tilde{p}=p+1$, and for all nonnegative integers $j_1,\ldots, j_{\tilde{p}}=0,1,2$ and $r=0,\ldots,4$, such that $j_1+\cdots+j_{\tilde{p}}+r\leq 6$, assume that the derivative 
	$$\frac{\partial^{j_1+\cdots+j_{\tilde{p}}+r} h_{in}^{\theta,\eta}}{\partial\theta_1^{j_1}\cdots\partial \theta_{\tilde{p}}^{j_{\tilde{p}}}\partial \eta^{r}}(y|x\, ,z),$$  
	exists for almost all $y$ and that 
	$$E_{0}\left( \sup_{i,\, n}\sup_{\theta\in \Theta} \sup_{g\in \mathcal{G}}\left \vert \frac{\partial^{j_1+\cdots+j_{\tilde{p}}+r} h_{in}^{\theta,\eta_i}}{\partial\theta_1^{j_1}\cdots\partial\theta_{\tilde{p}}^{j_{\tilde{p}}}\partial \eta^{r}}(Y_{in}|X_{in}\, , Z_{in})\right \vert^{2} \right)< \infty, \qquad \mathrm{with}\qquad \eta_i=g(Z_{in}).$$
	Assume that 
	\begin{equation}
	\sup_{z}\sup_{\theta}\sup_{\eta}\left \vert \frac{\partial^{j}}{\partial \theta^{j}}H^{(k)} ( \eta;\theta, z)\right \vert < \infty,
	\label{Assu}
	\end{equation}
	for $j=0,1,2$ and $k=2,3,4$ such that $j+k \leq 4,$ with 
	$$H^{(k)} ( \eta;\theta, z)=\frac{\partial^{k}}{\partial \eta^{k}}H ( \eta; \theta, z).$$ 
	Let 
	$$\widehat{H}(\eta;\theta,z)=\frac{\sum_{i=1}^{n}h_{in}^{\theta,\eta}(Y_{in}|X_{in},z)K_{in}(z)}{\sum_{i=1}^{n}K_{in}(z)};$$
	then, $\widehat{g}_{\theta}(z)$ is a solution of  $\widehat{H}^{(1)}(\eta;\theta,z)=0$
	with respect to $\eta$ for each fixed $\theta\in \Theta$ and $z\in \mathcal{Z}$. \\
	If we assume that Assumptions A1-A6 are satisfied, then we have, for all $j=0,1,2$,
	\begin{eqnarray}
	\sup_{\theta}\sup_{ z} \left \vert  \frac{\partial^{j}}{\partial \theta^{j}} \left (\widehat{g}_{\theta}(z)- g_{\theta}(z)\right )\right \vert = o_{p}(1).
	\label{NP1}
	\end{eqnarray}
\end{lemma} 
The assumptions used in the previous lemma are satisfied under the conditions used in the main results. \textbf{Condition I} is needed to ensure the identifiability of the arbitrary parameter $\theta_1$ (it plays the role of the true parameter $\theta_0$). This condition is verified when $\theta_1=\theta_0$ by the identifiability of our model (\ref{md1}). \textbf{Condition S} allows integrals to be interchanged with differentiation; this will be combined with the   implicit function theorem \citep[see][]{saaty2012nonlinear} to ensure the differentiability of  $\hat{g}_{\theta}(z)$ with respect to  $\theta $.  \\
Knowing that  $\Phi(\cdot)$ is a smooth function on  $\mathbb{R}$ and  $h_{in}^{\theta,\eta}(\cdot|\cdot \,,\cdot)$ is 
$$h_{in}^{\theta,\eta_i}(Y_{in}|X_{in} \,,Z_{in})=Y_{in}\log\left(\frac{\Phi(G_{in}(\theta,\eta_i))}{1-\Phi(G_{in}(\theta,\eta_i))}\right)-\log\left(1-\Phi(G_{in}(\theta,\eta_i))\right),$$
\textbf{Condition S} and Assumption (\ref{Assu})  are satisfied under the continuity condition of $\Phi(\cdot)$ and $\phi(\cdot)$, Assumption A9 and  the compactness of $\mathcal{X}$ and $\mathcal{Z}$. 
\subsection*{Proof of Lemma \ref{l4}}
The proof of this lemma is  similar to that of  Lemma~5 in \cite{severini1992profile}. Let us follow similar lines as in the proof of Lemma \ref{l8}  above, replacing  
$\zeta^{(j)}_{\theta}(Y_{in})$ by 
\begin{equation*}
\zeta_{\theta, \eta}^{(j,k)}(Y_{in},X_{in})=\frac{\partial ^{j}}{\partial \theta^{j}}\frac{\partial ^{k}}{\partial \eta^{k}} h_{in}^{\theta,\eta}(Y_{in}|X_{in} \,,z).
\end{equation*}
and 
Assumptions \textbf{H.1} and \textbf{H.2} in Lemma \ref{l8} by the following:
\begin{itemize}
	\item[\textbf{H.1'}] $\displaystyle\sup _{\theta}\sup_{\eta}\sup{ i, n} \left \vert\zeta_{\theta, \eta}^{(j,k)}(Y_{in},X_{in}) \right \vert < \infty, $ for $ j=0,\ldots,3, \; k=0,\ldots,5$ 
	\item[\textbf{H.2'}]  For all $k=0,\ldots,4,\; j=0,1,2$ and  $\theta\in \Theta,\; z \in \mathcal{Z}$, (\ref{dep1}) is satisfied and (\ref{dep2})  holds with $\zeta^{(j)}_{\theta}(Y_{in})$ replaced by $\zeta_{\theta, \eta}^{(j,k)}(Y_{in},X_{in})$.
\end{itemize}
Under the conditions used in  the lemma, it is clear that  \textbf{H.1'} is verified, and  \textbf{H.2'} is also satisfied by Assumption A3 (in particular, conditions (\ref{dependence1})).\\      
Using the results of Lemma~\ref{l8}, we have the following for all $ j=0,1,2$:
\begin{eqnarray}
\sup_{\theta,\, \eta,\, z} \left \vert\frac{\partial^{j}}{\partial \theta^{j}} \left ( \widehat{H}^{(1)}_n(\eta;\theta, z)-H^{(1)}(\eta;\theta, z)\right ) \right \vert=o_p(1),
\label{CEQ1}\\
\sup_{\theta,\, \eta,\, z} \left \vert \frac{\partial^{j}}{\partial \theta^{j}} \left ( \widehat{H}^{(2)}_n(\eta;\theta, z)-H^{(2)}(\eta;\theta, z)\right ) \right \vert=o_p(1),
\label{CEQ2}\\
\sup_{\theta,\, \eta,\, z} \left \vert \frac{\partial^{j}}{\partial \theta^{j}} \left ( \widehat{H}^{(3)}_n(\eta;\theta, z)-H^{(3)}(\eta;\theta, z)\right ) \right \vert=o_p(1),
\label{CEQ3}\\
\sup_{\theta,\, \eta,\, z} \left \vert \frac{\partial^{j}}{\partial \theta^{j}} \left ( \widehat{H}^{(4)}_n(\eta;\theta, z)-H^{(4)}(\eta;\theta, z)\right ) \right \vert=o_p(1).
\label{CEQ4}
\end{eqnarray}
Under Assumption A1,  for any $\epsilon >0$, there exists $\gamma>0$ such that
\begin{eqnarray*} 
	P\left(\sup_{\theta, z}\vert \widehat{g}_{\theta}(z)-g_{\theta}(z) \vert> \epsilon\right) 
	&\leq &P\left(\sup_{\theta, z} \vert H^{(1)}( \theta, \widehat{g}_{\theta}(z), z)\vert >\gamma\right)\\
	&=& P\left(\sup_{\theta, z} \vert \widehat{H}^{(1)}( \widehat{g}_{\theta}(z);\theta, z)- H^{(1)}( \widehat{g}_{\theta}(z);\theta, z)\vert >\gamma\right)\\
	&\leq &P\left(\sup_{\theta, z,\eta} \vert \widehat{H}^{(1)}(\eta;\theta, z)- H^{(1)}(\eta;\theta, z)\vert >\gamma\right).
\end{eqnarray*}
Hence, 
\begin{equation}
\sup_{\theta, z}\vert \widehat{g}_{\theta}(z)-g_{\theta}(z) \vert=o_{p}(1)
\label{EqNPR}
\end{equation}
The remainder of the proof is very similar  to that of Lemma 5 in \cite{severini1992profile} (p. 1798--1799); for the sake of completeness, we present the details.\\
We have by {\bf Condition I}
\begin{equation*}
\inf_{\theta}\inf_{z} -H^{(2)}( g_{\theta}(z);\theta,z)>0.
\end{equation*}
In addition, by {\bf Condition S}, for every $\delta>0$, there exists $\epsilon > 0$ such that  
\begin{equation*}
\sup_{\theta} \sup_{z} \sup_{\eta_1,\eta_2 : \vert \eta_1-\eta_2\vert\leq\epsilon} \left \vert H^{(2)}(\eta_2;\theta ,z)-H^{(2)}(\eta_1;\theta,z)\right \vert < \delta.  
\end{equation*} 
Hence, there exists $\epsilon > 0$ such that 
\begin{equation}
\inf_{\theta}\inf_{z} \inf_{\vert  \eta- g_{\theta}(z)\vert \leq \epsilon } \left \vert H^{(2)}(\eta;\theta,z)\right \vert >0.
\label{EQ1}
\end{equation}
Because $g_{\theta}(z)$ and $\widehat{g}_{\theta}(z)$ satisfy  
\begin{equation*}
H^{(1)}(g_\theta (z);\theta,z)= 0 \qquad \mbox{and} \qquad \widehat{H}^{(1)}(\widehat{g}_\theta (z);\theta,z)= 0, 
\end{equation*}
respectively, for each $\theta$ and $z$, it follows that 
\begin{eqnarray}
0&=& \widehat{H}^{(1)}(\widehat{g}_\theta (z);\theta,z) - H^{(1)}(g_\theta (z);\theta,z) \nonumber\\
&=&\widehat{H}^{(1)}(\widehat{g}_\theta (z);\theta,z) -H^{(1)}(\widehat{g}_\theta (z);\theta,z)+H^{(1)}(\widehat{g}_\theta (z);\theta,z) - H^{(1)}(g_\theta (z);\theta,z)\nonumber\\ 
&=&  r_n(\theta,z) +d_n(\theta, z) \left( \widehat{g}_{\theta}(z)-g_{\theta}(z)\right),
\label{EQ9}
\end{eqnarray}
for each $\theta, \; z$, where 
\begin{equation*}
r_n(\theta,z)=\widehat{H}^{(1)}(\widehat{g}_\theta (z);\theta,z) -H^{(1)}(\widehat{g}_\theta (z);\theta,z)\qquad \mbox{and } \;  	d_n(\theta,z)=\int_{0}^{1}H^{(2)}(tg_\theta (z)+(1-t)\widehat{g}_{\theta}(z);\theta,z)dt.
\end{equation*} 
Note that by (\ref{EQ1}) and  $\sup_{\theta}\Vert \widehat{g}_\theta -g_\theta\Vert =o_p(1)$, we have 
\begin{equation}
\lim\inf \inf_{z}\inf_{\theta} \left\vert \widehat{H}^{(2)}(\widehat{g}_\theta (z);\theta,z) \right\vert > 0  \qquad \mbox{and } \quad\lim\inf \inf_{z}\inf_{\theta} \left\vert d_n(\theta, z) \right\vert > 0 \qquad \mbox{as} \quad n\to \infty.
\label{EQ2}
\end{equation}
Because
\begin{equation*}
\widehat{H}^{(1)}(\widehat{g}_\theta (z);\theta,z)=0,
\end{equation*}
for all $\theta, \, z,$ we have
\begin{equation*}
\widehat{H}^{(2)}(\widehat{g}_\theta (z);\theta,z)\frac{\partial \widehat{g}_{\theta}}{\partial \theta}(z) + \frac{\partial \widehat{H}^{(1)}}{\partial \theta}(\widehat{g}_\theta (z);\theta,z)=0.
\end{equation*}
Then, we can deduce from (\ref{EQ2}), (\ref{CEQ1}), and (\ref{CEQ2}) that 
\begin{equation*}
\sup_{\theta}\sup_{z}\left \vert \frac{\partial \widehat{g}_{\theta}}{\partial \theta} (z) \right \vert =O_p(1).
\end{equation*} 
Similarly, we have
\begin{equation}
\sup_{\theta}\sup_{z}\left \vert \frac{\partial^j \widehat{g}_{\theta}}{\partial \theta^j} (z) \right \vert =O_p(1), \qquad\qquad j=0,1,2. 
\label{EQ7} 
\end{equation}  
Then, (\ref{EQ7}) and  (\ref{CEQ1})--(\ref{CEQ4}) yield
\begin{equation}
\sup_{\theta}\sup_{z}\left \vert \frac{\partial^j}{\partial \theta^j} r_n(\theta, z)\right \vert= o_p(1), \qquad \mbox{and}\qquad \sup_{\theta}\sup_{z}\left \vert \frac{\partial^j}{\partial \theta^j} d_n(\theta, z)\right \vert= O_p(1), \qquad\qquad j=0,1,2. 
\label{EQ8} 
\end{equation}
Now, differentiating (\ref{EQ9}) with respect to $\theta$ yields 
\begin{equation}
\frac{\partial r_n }{\partial \theta}(\theta, z)+ \left( \widehat{g}_\theta(z)-g_{\theta}(z)\right)\frac{\partial d_n }{\partial \theta}(\theta,z)+ d_n(\theta,z)\left( \frac{\partial \widehat{g}_\theta}{\partial \theta}(z)- \frac{\partial g_{\theta}}{\partial \theta}(z)\right)=0.
\end{equation}
Then, by (\ref{CEQ1})--(\ref{EQ8}),
\begin{equation*}
\sup_{\theta} \sup_{z}\left \vert \frac{\partial \widehat{g}_\theta}{\partial \theta}(z)- \frac{\partial g_{\theta}}{\partial \theta}(z)\right \vert =o_p(1).
\end{equation*}
On can similarly obtain 
\begin{equation*}
\sup_{\theta} \sup_{z}\left \vert \frac{\partial^2 \widehat{g}_\theta}{\partial \theta^2}(z)- \frac{\partial^2 g_{\theta}}{\partial \theta^2}(z)\right \vert =o_p(1).
\end{equation*}
This completes the proof. 	$\quad\square$  
\section*{Proof of Theorem \ref{th1}}
By Lemmas \ref{l1} and \ref{l2}, $Q_n$ converges to $Q$ in probability uniformly, i.e.,  
\begin{equation}
\sup_{\theta\in \Theta}\left \vert Q_n(\theta, g_\theta)-Q(\theta, g_\theta)\right \vert =o_p(1).
\label{UCQ}
\end{equation}
This result allows one to obtain
\begin{equation}\left \vert Q(\hat{\theta},g_{\hat{\theta}})-Q(\theta_{0},g_{0})\right \vert=o_p(1).
\label{UCQ2}
\end{equation}
Indeed, using $\vert \sup a - \sup b\vert \leq\sup \vert a-b\vert$, we have
\begin{eqnarray*}
	\left \vert Q(\hat{\theta},g_{\hat{\theta}})-Q(\theta_{0},g_{0})\right \vert &\leq& \left \vert Q_n(\hat{\theta}, \hat{g}_{\hat{\theta}})-Q(\hat{\theta},g_{\hat{\theta}})\right \vert + \left \vert Q_n(\hat{\theta}, \hat{g}_{\hat{\theta}})- Q(\theta_{0},g_{0})\right \vert \\
	& \leq &  \sup_{\theta} \left \vert Q_n(\theta, \hat{g}_{\theta})-Q(\theta,g_\theta)\right \vert +\left \vert \sup_{\theta} Q_n(\theta, \hat{g}_{\theta})- \sup_\theta  Q(\theta, g_\theta)\right \vert\\
	&\leq & 2 \sup_{\theta} \left \vert Q_n(\theta, \hat{g}_{\theta})-Q(\theta,g_\theta)\right \vert\\
	&\leq & 2 \sup_{\theta} \left \vert Q_n(\theta, \hat{g}_{\theta})-Q_n(\theta,g_\theta)\right \vert + 2 \sup_{\theta} \left \vert Q_n(\theta, g_\theta)-Q(\theta,g_\theta)\right \vert\\
	&=&o_p(1),
\end{eqnarray*}
by Lemma \ref{l3}, (\ref{UCQ}) and $\sup_\theta Q(\theta, g_\theta)=Q(\theta_0, g_0)$ (see Assumption A8). \\  

By Assumption A8, we have for a given $\theta\in \Theta$ that there exists $\varepsilon > 0$ and an open neighbourhood $N_{\theta}$ such that 
\begin{equation}\inf_{\theta_{1}\in N_{\theta}}\left\vert Q(\theta_{1},g_{\theta_{1}})-Q(\theta_{0},g_{0})\right \vert >\varepsilon.
\label{UCQ4}
\end{equation}
This and  (\ref{UCQ2}) imply that
\begin{equation}\mathbb{P}_{0}\left(\hat{\theta}\in N_{\theta}\right)\leq \mathbb{P}_{0}\left(\left\vert Q(\hat{\theta},g_{\hat{\theta}})-Q(\theta_{0},g_{0})\right \vert >\varepsilon\right)\rightarrow 0, \; \mathrm{as}\; n \to \infty.
\label{UCQ3}
\end{equation}

Let $N_{0}$ be an open neighbourhood of $\theta_{0}$, and consider the compact set $\Theta_{0}=\Theta\setminus N_{0}$. Let $\{N_{\theta}:\, \theta \in \Theta,\, \theta\neq \theta_{0}\}$ denote the open covering of $\Theta_{0}$ by the  procedure given above (each neighbourhood $N_\theta$ satisfies (\ref{UCQ4})). By the compactness of $\Theta_{0}$, let $\{N_{\theta_{1}},\ldots,N_{\theta_{r}}\}$ be a finite sub-covering; then, 
$$\mathbb{P}_{0}\left(\hat{\theta}\notin N_{0}\right)=\mathbb{P}_{0}\left(\hat{\theta}\in \Theta_{0}\right)\leq \sum_{j=1}^r \mathbb{P}_{0}\left(\hat{\theta}\in N_{\theta_{j}}\right)\rightarrow 0,\; \; \mathrm{as}\; n \to \infty,$$
by (\ref{UCQ3}).  
Therefore, we can conclude that 
$$\hat{\theta}-\theta_{0}=o_{p}(1), \qquad \mathrm{as}\qquad n\to \infty. $$
This yields the proof of Theorem~\ref{th1}.$\qquad \square$
\section*{Lemmas \ref{l1}-\ref{l3}}
We use the following notation:  
\begin{equation*}
\eta_{i}=g(Z_{in});\quad \tilde{U}_{in}=\tilde{U}_{in}(\theta,\eta_{i}); \qquad  \Phi_{in}=\Phi(G_{in}(\theta,g_{\theta}) ); \qquad \Lambda_{in}=\Lambda(G_{in}(\theta,g_{\theta})),    
\end{equation*}
for all  $\theta \in \Theta$, $1\leq i\leq n, \, n=1,2,\ldots$,  
with $\Lambda(\cdot)=\phi(\cdot)/\Phi(\cdot)(1-\Phi(\cdot))$.\\
The partial derivatives of  $S_{n}(\theta,g)$ with respect to $g$ of order $s=1,2,\ldots$, for any functions $v_{1}, \ldots, v_{s}$ in $\mathcal{G}$, are given  by
$$\frac{\partial^{s}S_{n}}{\partial g^{s}}(\theta,g)(v_{1},\cdots, v_{s})=n^{-1}\sum_{i=1}^{n}\xi_{in}\frac{\partial^{s}\tilde{U}_{in}}{\partial \eta^{s}}(\theta,\eta_{i})v_{1}(Z_{in})\cdots v_{s}(Z_{in}).$$

\begin{lemma}\label{l1} Under Assumptions A3, A6 and A9, we have
	for all $\theta \in \Theta $, 
	\begin{equation}
	S_{n}\left(\theta,g_{\theta}\right)-S\left(\theta,g_{\theta}\right)=o_{p}(1).
	\label{EL11}
	\end{equation}
	In addition, we have  
	\begin{equation}
	Q_{n}\left(\theta,g_{\theta}\right)-Q\left(\theta,g_{\theta}\right)=o_{p}(1),
	\label{EL12}
	\end{equation}
	if $M_n-M=o_p(1)$. 
\end{lemma}
Note that if Assumption A10 is satisfied, then $M_n-M=o_p(1)$.
\subsection*{Proof of Lemma \ref{l1}}
Let us  start with the  proof of (\ref{EL11}). We remark that
\begin{equation*}
S_n(\theta,g_\theta)=n^{-1}\xi_n^{T}\tilde{U}_{n}(\theta,g_\theta)=n^{-1}\sum_{i=1}^{n}\xi_{in}\tilde{U}_{in}(\theta,g_\theta),
\end{equation*}
where $\xi_{i}$ is the $q\times 1$ vector representing the  $i$th row in the matrix of instrumental variables.   
By definition (see (\ref{DefS})), we have $\mathbb{E}_0\left(S_n(\theta,g_\theta)\right)-S(\theta,g_\theta)=o(1)$. Then, it suffices to show that 
\begin{equation}
S_n(\theta,g_\theta)-\mathbb{E}_0\left( S_n(\theta,g_\theta)\right)=o_{p}(1).
\label{EL13}
\end{equation} 
Indeed  (omitting the $(\theta,g_\theta)-$arguments to simplify the notation), we have
\begin{eqnarray}
\mathbb{E}_0\left(\left\Vert S_n-\mathbb{E}_0\left( S_n\right)\right\Vert^{2}\right)&=&
n^{-2}\sum_{ i,j=1}^{n} \mathbb{E}_0\left(\left(\xi_{in}\tilde{U}_{in}- \mathbb{E}_0(\xi_{in}\tilde{U}_{in}) \right)^{T}  
\left(\xi_{jn} \tilde{U}_{jn}-\mathbb{E}_0(\xi_{jn}\tilde{U}_{jn})\right)\right) \nonumber \\
&\stackrel{( \ref{DepH})}{\leq}& n^{-2}\sum_{ i,j=1}^{n} \alpha_{ijn}\sum_{t=1}^{q}\left\{ \mbox{Var}_0\left( \xi_{itn}\tilde{U}_{in}\right) \mbox{Var}_0\left( \xi_{jtn}\tilde{U}_{jn}\right) \right\}^{1/2} \nonumber \\
&\leq& C n^{-2}\sum_{ i,j=1}^{n} \alpha_{ijn}=O\left(n^{-1}\sum_{s=1}^{\sqrt{n}}s\varphi(s)\right)=o(1),\nonumber
\label{SnES}
\end{eqnarray}
because $\mbox{Var}_{0}(\xi_{itn}\tilde{U}_{in})$ is bounded uniformly on $\theta$, $i$, and $t=1,\ldots,q$ (by  Assumption A6) and because $\varphi(s)\to $ as $s\to +\infty$ (by assumption  A3). This completes the proof of (\ref{EL13}) and thus that of (\ref{EL11}).\\ The proof of (\ref{EL12}) is made straightforward by combining (\ref{EL11}) with Assumption A10.$\qquad \square$

\begin{lemma}\label{l2}   Under Assumptions A6-A9, we have  
	$S_{n}\left(\cdot,g_{\cdot}\right)-S\left(\cdot,g_{\cdot}\right)$ is stochastically equicontinuous on $\Theta$. \\ In addition, if $M_n-M=o_p(1)$, then  we have 
	$Q_{n}\left(\cdot,g_{\cdot}\right)-Q\left(\cdot,g_{\cdot}\right)$ is also stochastically equicontinuous on $\Theta$.  
\end{lemma}
\subsection*{Proof of Lemma~\ref{l2}}
Stochastic equicontinuity in $\Theta$ can be obtained by proving that $S_{n}(\theta,g_{\theta})$ satisfies  a stochastic Lipschitz-type condition on $\theta$  \citep[see][p.~17]{matyas1999generalized}.\\
Let us show that $S_{n}(\cdot,g_\cdot)$ is stochastically equicontinuous on $\theta$ because $S(\cdot,g_\cdot)$ is continuous by Assumption A8. It suffices to show that \citep{andrews1992generic} for each $\theta_1, \theta_2 \in \Theta$:
\begin{equation}
\left\Vert S_{n}(\theta_1, g_{\theta_1})-S_{n}(\theta_2 ,g_{\theta_2})\right \Vert=O_{p} \left(\Vert \theta_1-\theta_2\Vert\right).
\label{EquiCon}
\end{equation}
Indeed, for $\theta_1, \theta_2 \in \Theta$,
\begin{eqnarray*}
	\left\Vert S_{n}(\theta_1, g_{\theta_1})-S_{n}(\theta_2 ,g_{\theta_2})\right \Vert &\leq & n^{-1} \sup_{i , \, n} \Vert \xi_{in}\Vert \sum_{i=1}^{n}  \left \vert \tilde{U}_{in}(\theta_1, g_{\theta_1})-\tilde{U}_{in}(\theta_2, g_{\theta_2}) \right\vert \\
	&\leq & n^{-1} \sup_{i , \, n} \Vert \xi_{in}\Vert  \sum_{i=1}^{n}  \left\{ \sup_{\theta,\, \eta} \left\Vert \frac{\partial\tilde{U}_{in}}{\partial\theta}(\theta,\eta)\right \Vert \Vert \theta_1-\theta_2\Vert\right. \\
	&\,& \qquad \qquad\qquad +\left . \sup_{\theta,\, \eta} \left\vert \frac{\partial\tilde{U}_{in}}{\partial\eta}(\theta,\eta)\right \vert \Vert g_{\theta_1}-g_{\theta_2}\Vert\right\}\\
	&\leq & n^{-1} \sup_{ i , \, n} \Vert \xi_{in}\Vert \sum_{i=1}^{n}\left\{ \sup_{\theta,\, \eta} \left\Vert \frac{\partial\tilde{U}_{in}}{\partial\theta}(\theta,\eta)\right \Vert \right.\\
	&\,& \qquad \qquad\qquad +\left . \sup_{\theta}\left\Vert \frac{\partial g_{\theta}}{\partial \theta}\right\Vert  \sup_{\theta,\, \eta} \left\vert \frac{\partial\tilde{U}_{in}}{\partial\eta}(\theta,\eta)\right \vert \right\}\Vert \theta_1-\theta_2\Vert.
\end{eqnarray*}
By Assumption A6 and  Proposition~\ref{prop1}, we have that $\sup_{i , \, n} \Vert \xi_{in}\Vert$ is bounded and $\sup_{\theta}\left\Vert\frac{\partial g_{\theta}}{\partial \theta}\right \Vert$ is finite, respectively. Then, we have to show that
\begin{equation}
n^{-1} \sum_{i=1}^{n}  \sup_{\theta, \eta} \left\Vert \frac{\partial\tilde{U}_{in}}{\partial\theta}(\theta,\eta)\right \Vert +  \sup_{\theta, \eta} \left\vert \frac{\partial\tilde{U}_{in}}{\partial\eta}(\theta,\eta)\right \vert =O_{p}(1);
\end{equation}
This is  equivalent to
\begin{equation}
\sup_{\theta, \eta} \left\Vert \frac{\partial\tilde{U}_{in}}{\partial\theta}(\theta,\eta)\right \Vert =O_{p}(1), \qquad 1\leq i\leq n , \, n=1,2,\ldots
\label{U1Borne}
\end{equation}
and 
\begin{equation}
\sup_{\theta, \eta} \left\vert \frac{\partial\tilde{U}_{in}}{\partial\eta}(\theta,\eta)\right \vert=O_{p}(1), \qquad 1\leq i\leq n , \, n=1,2,\ldots
\label{U2Borne}
\end{equation}
Let us  prove (\ref{U1Borne}) in the following. The proof of (\ref{U2Borne}) follows the same lines and is thus omitted.\\

\noindent 
\underline{\textbf{Proof of (\ref{U1Borne}):}}
\\
Recall that $$\Lambda(t)=\frac{\phi(t)}{\Phi(t)(1-\Phi(t))}.$$ By definition, we have
\begin{equation*}
\tilde{U}_{in}(\theta, \eta)=\Lambda(G_{in}(\theta, \eta))\left(Y_{in} -\Phi(G_{in}(\theta, \eta))\right),
\end{equation*}
with  $G_{in}(\theta, \eta)=a_{in}(\theta)b_{in}(\theta,\eta)$,
where $a_{in}(\cdot)$ and $b_{in}(\cdot)$ are defined  by 
\begin{equation}
a_{in}(\theta):=(v_{in}(\lambda))^{-1}\qquad \mathrm{and } \qquad b_{in}(\theta, \eta):=X_{in}^{T}\beta+\eta, \qquad 1\leq i\leq n , \, n=1,2,\ldots,
\end{equation}
with $\theta^{T}=(\beta^{T}, \lambda)$. We have 
\begin{eqnarray}
\frac{\partial \tilde{U}_{in}}{\partial \theta}(\theta, \eta)&=& \left\{\Lambda^{'}(G_{in}(\theta, \eta))(Y_{in}-\Phi(G_{in}(\theta, \eta)))\right. \nonumber \\ & & \qquad \qquad - \left.\Lambda(G_{in}(\theta, \eta))\phi(G_{in}(\theta, \eta)) \right\} \frac{\partial G_{in}}{\partial\theta}(\theta, \eta)
\end{eqnarray}
where $\Lambda^{'}(\cdot)$ denotes the derivative of  $\Lambda(\cdot)$.


Let us first establish that 
\begin{equation}
\sup_{t\in \mathcal{M}, y\in \{0,1\}} \left \vert \Lambda^{'}(t)(y-\Phi(t))-\phi(t)\Lambda(t)\right \vert < \infty, 
\label{boundedU}
\end{equation}
which is equivalent to showing that  $\Lambda^{'}(t)$ and $\phi(t)\Lambda(t)$ are bounded uniformly in $t\in \mathcal{M}$ (the definition of $\mathcal{M}$ is given in {\bf A.1}). Because $\phi^{'}(t)=-t\phi(t)$, we can rewrite $\Lambda^{'}(t)$ as 
\begin{equation}
\Lambda^{'}(t)=\frac{1}{\Phi(t)}\left\{\frac{\phi(t)}{1-\Phi(t)}\left( \frac{\phi(t)}{1-\Phi(t)} -t\right)\right\} - \frac{\phi^{2}(t)}{\Phi^{2}(t)(1-\Phi(t))}.
\label{lambdstr}
\end{equation}
Notice that $\Lambda(\cdot)$ and $\Lambda^{'}(\cdot)$ may be unbounded only  at $\pm \infty$,  and because $\mathcal{M}$ is a compact subset of $\mathbb{R}$, these functions are bounded on $\mathbb{R}$. This establishes  (\ref{boundedU}). \\
\noindent 
We remark that
\begin{equation}
\left \Vert\frac{\partial G_{in}(\theta, \eta)}{\partial\theta}\right \Vert\leq\left \Vert \frac{\partial a_{in} (\theta)}{\partial\theta}\right \Vert  \left \vert b_{in}(\theta, \eta) \right\vert +\left \Vert \frac{\partial b_{in}(\theta,\eta)} {\partial\theta}  \right \Vert \left \vert a_{in}(\theta) \right\vert.
\label{Githeta}
\end{equation}
Then, $\left \Vert\frac{\partial G_{in}(\theta, \eta)}{\partial\theta}\right \Vert$ is bounded uniformly in $i, n ,\theta, \eta$ by  Assumptions A6 and A9 and the compactness  of $\Theta$ (see assumption A7). This completes the proof of (\ref{U1Borne}); hence,  (\ref{EquiCon}) is proved.   $\qquad \square$

\begin{lemma}\label{l3} Under the assumptions of Proposition~\ref{prop1} and Assumptions A6 and A9, we have  
	\begin{equation}
	\sup_{\theta\in \Theta}\left\Vert S_{n}(\theta,\hat{g}_{\theta})-S_{n}(\theta,g_{\theta}) \right\Vert=o_{p}(1).
	\label{EL21}
	\end{equation}
	If in addition $M_n-M=o_p(1)$, then  we have  
	\begin{equation}
	\sup_{\theta\in \Theta}\left\vert Q_{n}(\theta,\hat{g}_{\theta})-Q_{n}(\theta,g_{\theta}) \right\vert=o_{p}(1).
	\label{EL22}
	\end{equation}
\end{lemma}
\subsection*{Proof of Lemma \ref{l3}}
Let us prove (\ref{EL21}). For each $\theta \in \Theta$
\begin{eqnarray*}
	\left\Vert S_{n}(\theta,\hat{g}_{\theta})-S_{n}(\theta, g_{\theta})\right\Vert &=& n^{-1}\left\Vert \sum_{i=1}^{n} \xi_{i}\left(\tilde{U}_{in}(\theta,\hat{g}_{\theta})-\tilde{U}_{in}(\theta, g_{\theta})\right)\right\Vert\\
	&\leq & n^{-1} \sum_{i=1}^{n}\sup_{i,n}\left\Vert  \xi_{in} \right\Vert \left\vert \tilde{U}_{in}(\theta,\hat{g}_{\theta})-\tilde{U}_{i}(\theta, g_{\theta})\right\vert \\  
	&\leq & n^{-1} \sum_{i=1}^{n}\sup_{i, n}\left\Vert  \xi_{in} \right\Vert\sup_{\theta, \eta}\left\vert   \frac{\partial\tilde{U}_{in}}{\partial\eta}(\theta,\eta)\right\vert \sup_{\theta}\Vert \hat{g}_{\theta}-g_{\theta}\Vert\\ 
	&=& o_p(1),
\end{eqnarray*}
because $\sup_{i, n}\left\Vert  \xi_{in} \right\Vert= O_p(1)$ (by Assumption A6), $\sup_{\theta}\Vert \hat{g}_{\theta}-g_{\theta}\Vert=o_p(1)$ (see Proposition~\ref{prop1})  and $\sup_{\theta, \eta}\left\vert   \frac{\partial\tilde{U}_{in}}{\partial\eta}(\theta,\eta)\right\vert =O_p(1)$ uniformly on $i$ and $n$ (see the proof of Lemma~\ref{l2}).\\ 
The proof of  (\ref{EL22}) is made trivial by combining  (\ref{EL21}) with Assumption  A10. $\qquad \square$

\section*{Proof of Theorem \ref{th2}}
Recall that $\frac{d}{d\theta}Q_n(\theta,g_\theta)$ denotes differentiation with respect to $\theta$, while $\frac{\partial}{\partial \theta}Q_n(\theta,g_\theta)$ denotes  the partial derivative with respect to $\theta$.\\
Using a Taylor's series expansion and the fact that 
\begin{equation*}
\left.\frac{d}{d\theta} Q_{n} (\theta, \hat{g}_{\theta})\right\vert _{\theta=\hat{\theta}}=0, 
\end{equation*} 
we have  
\begin{eqnarray}
\hat{\theta}-\theta_{0}=-\left\{\left.\frac{d^{2}}{d\theta d\theta^T}Q_{n}(\theta,\hat{g}_{\theta})\right\vert_{\theta=\theta^{*}}\right\}^{-1}\left\{ \left.\frac{d }{d \theta}Q_{n} (\theta, \hat{g}_{\theta})\right\vert_{\theta=\theta_{0}}\right\},
\label{EQ10}
\end{eqnarray}
for some  $\theta^{*}$ between $\theta_0$ and $\hat{\theta}$.\\ 
First, we would like to replace $\hat{g}_{\theta}(.)$ in (\ref{EQ10}) with $g_{\theta}(.)$. For this,  let us show that  $\frac{d}{d\theta}Q_{n}(\theta,\hat{g}_{\theta})$ (resp. $\displaystyle \frac{d^2}{d\theta d\theta^T}Q_{n}(\theta,\hat{g}_{\theta})$) and $\frac{d}{d\theta}Q_{n}(\theta,g_{\theta})$ (resp. $\displaystyle\frac{d^2}{d\theta d\theta^T}Q_{n}(\theta,g_{\theta})$) have the same behavior as a function of $\theta$ in a neighbour of $\theta_{0}$. In other words,  
\begin{eqnarray}
\sup_{\theta} \left\Vert \frac{d^{2}}{d\theta d \theta^T}Q_{n}(\theta,\hat{g}_{\theta})-\frac{d^{2}}{d\theta d \theta^T}Q_{n}(\theta, g_{\theta})\right\Vert =o_{p}(1)
\label{TH22}
\end{eqnarray}
and 
\begin{eqnarray}
\left.\frac{d }{d \theta}Q_{n} (\theta, \hat{g}_{\theta})\right\vert _{\theta=\theta_0} -\left.\frac{d }{d \theta} Q_{n}(\theta, g_{\theta})\right\vert _{\theta=\theta_0}=o_{p}(1).
\label{TH21}
\end{eqnarray}

We remark that  (\ref{TH22}) is equivalent to 
\begin{equation}
\sup_{\theta} \left\Vert \frac{d}{d\theta}S_{n}(\theta,\hat{g}_{\theta})-\frac{d}{d\theta}S_{n}(\theta, g_{\theta})\right\Vert =o_{p}(1)
\label{TH221}
\end{equation}
and 
\begin{equation}
\sup_{\theta} \left\Vert \frac{d^2}{d\theta d\theta^T}S_{n}(\theta,\hat{g}_{\theta})-\frac{d^2}{d\theta d\theta^T}S_{n}(\theta, g_{\theta})\right\Vert =o_{p}(1)
\label{TH222}
\end{equation}
by (\ref{gmm}) (because $M_n-M=o_p(1)$ thanks to Assumption A10)
and
\begin{equation*}
\sup_{\theta} \left\Vert S_{n}(\theta,\hat{g}_{\theta})-S_{n}(\theta, g_{\theta})\right\Vert =o_{p}(1)
\label{TH223}
\end{equation*} 
(see Lemma~\ref{l3}).
Then,  (\ref{TH221}) and (\ref{TH222})  follow immediately from Lemma~\ref{l6}.\\ 
To prove (\ref{TH21}),  we have the following Taylor expansion 
$$\displaystyle \frac{d}{d\theta}\left( Q_{n}(\theta,\hat{g}_{\theta})-Q_{n}(\theta,g_{\theta})\right)=\frac{d}{d\theta} \left( \frac{\partial Q_{n}}{\partial g}(\theta,g_{\theta})(\hat{g}_{\theta}-g_\theta)+ \tilde{r}_n(\theta) \right),$$
where 
\begin{equation*}
\tilde{r}_n(\theta)=\int_{0}^1 \frac{\partial^2 Q_n}{\partial g^2}(\theta, g_\theta+t(\hat{g}_\theta-g_\theta))(\hat{g}_\theta- g_\theta)^2\,dt.
\end{equation*}
We have
\begin{equation*}
\left .\frac{d}{d\theta} \tilde{r}_n(\theta)\right \vert _{\theta=\theta_0}=o_p(1),
\end{equation*}
using similar arguments as for the terms  $\displaystyle\frac{d^j}{d\theta^j}r_n^{(1)}(\theta)$ for $j=0,1$ and $\displaystyle \frac{d^2}{d\theta d\theta^T}r_n^{(1)}(\theta)$ in Lemma~\ref{l6} below (see (\ref{EQl51})).
Therefore, we obtain
\begin{eqnarray*}
	\left.\frac{d }{d \theta}Q_{n} (\theta, \hat{g}_{\theta})\right\vert_{\theta=\theta_{0}}-\left.\frac{d }{d \theta}Q_{n} (\theta, g_{\theta})\right\vert_{\theta=\theta_{0}} 
	&=&\left.\frac{d }{d \theta}\frac{\partial Q_{n} }{\partial g} (\theta, g_{\theta})\right\vert_{\theta=\theta_{0}} (\hat{g}_0-g_0) \\
	&&\qquad + \frac{\partial Q_{n} }{\partial g} (\theta_0, g_{0}) (\hat{g}^{'}_0-g_0^{'}) +\left. \frac{d}{d\theta} r_{n}(\theta)\right \vert _{\theta=\theta_0},\\ 
	&=&o_p(1)
\end{eqnarray*} 
by Lemma~\ref{l5}, where $\displaystyle g^{'}_{0}(.)=\left.\frac{g_\theta}{\partial \theta^T}(.)\right\vert_{\theta=\theta_0}$.
\\
Consequently, we obtain 
\begin{eqnarray}
\hat{\theta}-\theta_{0}=-\left\{\left.\frac{d^{2}}{d\theta d\theta^T}Q_{n}(\theta,g_{\theta})\right\vert_{\theta=\theta^{*}}\right\}^{-1}\left\{ \left.\frac{d }{d \theta}Q_{n} (\theta, g_{\theta})\right\vert_{\theta=\theta_{0}}\right\}+o_p(1)
\label{EQ11}
\end{eqnarray}	
where $\theta^{*}$ is between $\hat{\theta}$ and $\theta_0$. \\	
{Let us show that for each $\theta^{*}$ lying between $\theta_0$ and $\hat{\theta}$,} 
\begin{equation*}
\left.\frac{d^{2}}{d\theta d\theta^T}Q_{n}(\theta,g_{\theta})\right \vert_{\theta=\theta^*}=2\,B_2(\theta_0)+o_p(1),
\end{equation*}
to replace the Hessian matrix in the right-hand side of (\ref{EQ11}) by its limit $B_2(\theta_0)$.\\ Let us consider the first- and second-order differentials of  $Q_n(\theta, g_\theta)$ with respect to $\theta$: 
\begin{equation}
\frac{d}{d\theta}Q_n(\theta,g_\theta)=2S_n^{T}(\theta,g_\theta)M_n\left\{\frac{\partial S_n}{\partial \theta}(\theta,g_\theta)+\frac{\partial S_n}{\partial g}(\theta,g_\theta)g^{'}_\theta\right\}
\end{equation}	
with $g^{'}_\theta$ being a $1\times\tilde{p}$ ($\tilde{p}=p+1$) matrix  given by   $\displaystyle \frac{\partial g_{\theta}}{\partial \theta^T}$ and 
\begin{eqnarray}
\frac{d^{2}}{d\theta d\theta^T}Q_{n}(\theta,g_{\theta})&=& 2\left\{\frac{\partial S_n}{\partial \theta}(\theta,g_\theta)+\frac{\partial S_n}{\partial g}(\theta,g_\theta)g^{'}_\theta\right\}^{T}M_n\left\{\frac{\partial S_n}{\partial \theta}(\theta,g_\theta)+\frac{\partial S_n}{\partial g}(\theta,g_\theta)g^{'}_\theta\right\} \nonumber \\
&& \qquad \qquad +  2S_n^{T}(\theta,g_\theta)M_n\frac{d}{d\theta^T}\left\{\frac{\partial S_n}{\partial \theta}(\theta,g_\theta)+\frac{\partial S_n}{\partial g}(\theta,g_\theta)g^{'}_\theta\right\}
\label{2Q}
\end{eqnarray}
with 
\begin{equation*}
\frac{d}{d\theta^T}\frac{\partial S_n}{\partial \theta}(\theta,g_\theta)= \frac{\partial^2 S_n}{\partial \theta\partial \theta^T}(\theta,g_\theta)+\frac{\partial^2 S_n}{\partial \theta\partial g}(\theta,g_\theta)g^{'}_\theta,
\end{equation*}
\begin{equation*}
\frac{d}{d\theta^T}\frac{\partial S_n}{\partial g}(\theta,g_\theta)= \frac{\partial^2 S_n}{\partial \theta\partial g}(\theta,g_\theta)+\frac{\partial^2 S_n}{\partial g^2}(\theta,g_\theta)\frac{\partial g_\theta}{\partial \theta}.
\end{equation*}
Note that 
$$S_n(\theta^*,g_{\theta^*})=S_n(\theta^*,g_{\theta^*})-S_n(\theta_0,g_{0})+S_n(\theta_0,g_{0})-S(\theta_0,g_0)=o_p(1),$$ 
because $S(\theta_0,g_0)=0$ and by Lemmas~\ref{l1}-\ref{l2}, 
$$S_n(\theta_0,g_{0})-S(\theta_0,g_{0})=o_p(1),$$ 
and because $\theta^*$ lies between $\hat{\theta}$ and $\theta_0$, by Lemma~\ref{l2}
$$S_n(\theta^*,g_{\theta^*})-S_n(\theta_0,g_{0})=o_p(1).$$
Using similar arguments as in the proof of (\ref{U1Borne}) in Lemma~\ref{l2} using Assumption A9 to ensure the boundedness when differentiating twice with respect to $\theta$, we have
\begin{equation}
\left\Vert \frac{d}{d\theta^T}\frac{\partial S_n}{\partial \theta}(\theta,g_\theta)\right\Vert =O_p(1)\qquad \mathrm{and } \qquad \left\Vert\frac{d}{d\theta^T}\frac{\partial S_n}{\partial g}(\theta,g_\theta)g^{'}_\theta\right\Vert=O_p(1).
\end{equation}
Then, we can ignore the second term in the right-hand side of (\ref{2Q}) at $\theta=\theta^*$. Hence, by Lemma~\ref{l7} and $\theta^*-\theta_0=o_p(1)$ (thanks to Theorem~\ref{th1}), we have 
\begin{equation*}
\frac{\partial S_n}{\partial \theta}(\theta^*,g_{\theta^*})-\frac{\partial S}{\partial \theta}(\theta_0,g_0)=o_p(1)
\end{equation*}
and 
\begin{equation*}
\frac{\partial S_n}{\partial g}(\theta^*,g_{\theta^*})g^{'}_{\theta^*}-\frac{\partial S}{\partial g}(\theta_0,g_{0})g^{'}_{0}=o_p(1),
\end{equation*}
with $\displaystyle g^{'}_{\theta^*}=\left.\frac{g_\theta}{\partial \theta^T}\right\vert_{\theta=\theta^{*}}$.\\
In addition, if   $M_n-M=o_p(1)$, we deduce  that
\begin{eqnarray*}
	\left.\frac{d^{2}}{d\theta d\theta^T}Q_{n}(\theta,g_{\theta})\right \vert_{\theta=\theta^*}&=& 2\, 
	\left\{\frac{\partial S}{\partial \theta}(\theta_0,g_{0})+\frac{\partial S}{\partial g}(\theta_0,g_{0})g^{'}_{0}\right\}^{T}M\left\{\frac{\partial S}{\partial \theta}(\theta_0,g_{0})+\frac{\partial S}{\partial g}(\theta_0,g_{0})g^{'}_{0}\right\}+o_p(1)\\
	&=& 2\,B_2(\theta_0)+o_p(1).
\end{eqnarray*}	
We remark that 
\begin{equation*}
\left.\frac{d}{d\theta}Q_n(\theta,g_\theta)\right\vert_{\theta=\theta_0}=2S_n^{T}(\theta_0,g_0)M_n\left\{\frac{\partial S_n}{\partial \theta}(\theta_0,g_0)+\frac{\partial S_n}{\partial g}(\theta_0,g_0)g^{'}_0\right\}.
\end{equation*}
Then, by (\ref{EqL2}) (see the proof of Lemma~\ref{l7}), we have
$$\frac{\partial S_n}{\partial \theta}(\theta_0,g_0)-\frac{\partial S}{\partial \theta}(\theta_0,g_0)=o_p(1)\qquad \mathrm{and } \qquad \frac{\partial S_n}{\partial g}(\theta_0,g_0)g^{'}_0-\frac{\partial S}{\partial g}(\theta_0,g_0)g^{'}_0=o_p(1).$$
Consequently, we obtain
\begin{equation*}
\left.\frac{d}{d\theta}Q_n(\theta,g_\theta)\right\vert_{\theta=\theta_0}=2S_n^{T}(\theta_0,g_0)M\left\{\frac{\partial S}{\partial \theta}(\theta_0,g_0)+\frac{\partial S}{\partial g}(\theta_0,g_0)g^{'}_0\right\}+o_p(1).
\end{equation*}
Then, we have 
\begin{equation*}
\hat{\theta}-\theta_{0}=-\left\{B_2(\theta_0)\right\}^{-1}\left\{\frac{\partial S}{\partial \theta}(\theta_0,g_0)+\frac{\partial S}{\partial g}(\theta_0,g_0)g^{'}_0\right\}^{T}M\,S_n(\theta_0,g_0)+o_p(1).
\end{equation*}
To end the proof, it remains to be shown that 
\begin{equation*}
\sqrt{n}B_{1}(\theta_0)^{-1/2}S_n(\theta_0,g_0)\longrightarrow \mathcal{N}(0,\mathbb{I}_{q}).
\end{equation*}
Consider, for all  $w\in \mathbb{R}^q$ such that $\Vert w\Vert=1$,  
\begin{eqnarray*}
	A_n &=&w^T\left\{\mathbb{E}_0\left(nS_n(\theta_0,g_0)S_n^T(\theta_0,g_0)\right)\right\}^{-1/2}\sqrt{n}S_n(\theta_0,g_0) \\
	&=& n^{-1/2}\sum_{i=1}^{n}B_{in},
\end{eqnarray*}
with
\begin{equation*}
B_{in}=w^T\left\{\mathbb{E}_0\left(nS_n(\theta_0,g_0)S_n^T(\theta_0,g_0)\right)\right\}^{-1/2}\xi_{in}\tilde{U}_{in}(\theta_0,g_0).
\end{equation*} 
By the Cramer-Wold device, it suffices to show that $A_n$ converges asymptotically to a standard normal distribution, for all $w\in \mathbb{R}^q$, such that $\Vert w\Vert=1$. \\
To prove this, we will use the central theorem limit (CTL) proposed by \cite{pinkse2007central}. These authors used an idea of \cite{bernstein1927extension} based on partitioning the observations into $J$ groups $\mathcal{G}_{n1},\ldots,\mathcal{G}_{nJ}$, $1\leq J< \infty$, which are divided up into mutually exclusive subgroups $\mathcal{G}_{j1n},\ldots,\mathcal{G}_{jm_{jn}n}$, $j=1,\ldots,J $. Each observation belongs to one subgroup, and its membership can vary  with the sample size $n$, as can the number of subgroups $m_{jn}$ in group $j$. We assume that the partition is constructed such that 
$$m_{jn}/m_{1n}=o(1)\qquad j=2,\ldots,J$$
and 
$$\mathrm{Card}(\mathcal{G}_{irn})=O\left(\mathrm{Card}(\mathcal{G}_{jtn})\right), \qquad \forall \, i,j=1,\ldots,J,\; r=1,\ldots, m_{in}\; , t=1,\ldots, m_{jn}.$$
Partial sums over elements in groups and subgroups are denoted by $A_{nj}$ and $A_{jtn}$,$j=1,\ldots,J$, and $t=1,\ldots,m_{jn}$, respectively. Thus, we have 
$$A_n=\sum_{j=1}^{J}A_{jn}=\sum_{j=1}^{J}\sum_{t=1}^{m_{jn}}A_{jtn},\qquad A_{jtn}=n^{-1/2}\sum_{i\in\mathcal{G}_{jtn}}B_{in}.$$
Let us recall in the following the assumptions under which the CTL of \cite{pinkse2007central} holds.\\
\textbf{Assumption A.} For any $j=1,\ldots,J$, let $\mathcal{G}^{*},\, \mathcal{G}^{**} \subset \mathcal{G}_{jn}$ be any sets for which $$\forall t=1,\ldots, m_{jn} \, :\; \mathcal{G}^{*}\cap \mathcal{G}_{jtn}\neq\emptyset \qquad \Rightarrow \quad \mathcal{G}^{**}\cap \mathcal{G}_{jtn}=\emptyset.$$  
Then, for any function $f$ in $\mathcal{F}=\left\{f: \forall t\in \mathbb{R} f(t)=t \; \mathrm{or} \; \exists \upsilon \in \mathbb{R} : \forall t\in \mathbb{R} f(t)=e^{\iota \upsilon t}\right\}$, where $\iota$ is the imaginary number 
\begin{align*}
&\left\vert \mathrm{Cov}\left( f\left(\sum_{i\in \mathcal{G}^*}B_{in}\right),f\left(\sum_{i\in \mathcal{G}^{**}}B_{in}\right)\right)\right\vert \leq \nonumber \\&  \qquad \qquad\qquad \left\{ \mathrm{Var}\left( f\left(\sum_{i\in \mathcal{G}^*}B_{in}\right)\right)\mathrm{Var}\left( f\left(\sum_{i\in \mathcal{G}^{**}}B_{in}\right)\right)\right\}^{1/2}\alpha_{jn},
\label{A1}
\end{align*}
for some mixing numbers $\alpha_{jn}$ with 
\begin{equation*}
\lim_{n\to \infty} \sum_{j=1}^{J}m_{jn}^2\alpha_{jn}=0.
\label{A2}
\end{equation*} 
\textbf{Assumption B.} 
$$\lim_{n\to \infty }\max_{t\leq m_{jn}} \frac{\sigma_{jtn}}{\gamma_{jn}}=0,\; j=1,\ldots,J,\qquad \lim_{n\to \infty } \frac{\gamma_{jn}}{\gamma_{1n}} =0,\; j=2,\ldots,J,$$
where 
$$\sigma_{jtn}^2=\mathbb{E}_{0}(A_{jtn}^2),\qquad \mathrm{and}\qquad \gamma_{nj}^2= \sum_{t=1}^{m_{jn}}\sigma_{jtn}^2.$$
\textbf{Assumption C.} For some $\tau>1$
$$\mathbb{E}_0\left(\vert A_{jtn}\vert^{2\tau} \right)=o\left(\sigma_{jtn}^2 \gamma_{jn}^{2\tau-2}\right), \; j=1,\ldots,J,\; t=1,\ldots, m_{jn}.$$
If assumptions $A-C$ hold, then by Theorem~1 in \cite{pinkse2007central}, we have $A_n\longrightarrow \mathcal{N}(0,1).$ Thus, to complete the proof, we have to check these assumptions in our context.\\

\noindent\underline{\textbf{Assumption A}:} This holds under (\ref{DepH}) (Assumption A3).\\
Let us choose for instance $J = 2$ groups, each with $m_{1n}, m_{2n}$ subgroups such that $m_{2n}=o(m_{1n})$. Each subgroup is viewed as an area of size $O(\sqrt{c_n}\times \sqrt{c_n})$ such that $(m_{1n}+m_{2n})c_n=O(n)$. Because $\varphi(\cdot)$ is a decreasing function (Assumption A3), $\alpha_{jn}=O(\varphi(\sqrt{c_n})) $ for $j=1,2$. The sequence $c_n$ must be such that 
$c_n=O(n^{-\nu+1/2})$ for some $0<\nu < 1/2$ and  $n^{\nu +1/2} \varphi(\sqrt{c_n}) \to 0 $ as $n\to \infty $.\\
If for instance $\varphi(t)=O(t^{-\iota})$, then $n^{\nu +1/2} \varphi(\sqrt{c_n})= O(n^{\iota (\nu -1/4)+(1+\nu)/2})$; this tends to $0$ for each $\iota>2(1+\nu)/(1-4\nu)$.\\ 
\underline{\textbf{Assumption B} : }  By assumption A10, $B_1(\theta_0)$ is positive definite and by definition is the limit of \\ $\mathbb{E}_0\left(nS_n(\theta_0,g_0)S^T_n(\theta_0,g_0)\right)$. Then, for sufficiently large $n$, the last matrix is positive definite, and its inverse is $O(1)$. Therefore, $B_{in}$ is bounded uniformly on $i$ and $n$ because $\xi_{in}$ is bounded uniformly on $i$ and $n$ by Assumption A6, as is $\tilde{U}_{in}(\theta_0,g_0)$.  
Then, for all $j=1,\ldots,J$ and $t=1,\ldots, m_{nj}$,
$$\sigma_{jtn}=\left\{n^{-1}\mathbb{E}_0\left(\sum_{i \in \mathcal{G}_{jtn}}B_{in}\right)\right\}^{1/2}=O\left(n^{-1/2}\mathrm{Card}(\mathcal{G}_{jtn})\right)$$
and 
$$\gamma_{jn}=O\left(\frac{m_{jn}}{\sqrt{n}}\max_{t\leq m_{jn}}\mathrm{Card}(\mathcal{G}_{jtn})\right).$$
Therefore,
$$\frac{\sigma_{jtn}}{\gamma_{jn}}=O(1/m_{jn})\to 0 \; \mathrm{as} \; n\to \infty,  $$
for all $j=1,\ldots,J$ and $t=1,\ldots, m_{jn}$.\\
Now, consider  the second limit in Assumption B. We have for all $j=2,\ldots,J$
$$\frac{\gamma_{jn}}{\gamma_{1n}}=O\left(\frac{m_{jn}\max_{t\leq m_{jn}}\mathrm{Card}(\mathcal{G}_{jtn})}{m_{1n}\max_{t\leq m_{1n}}\mathrm{Card}(\mathcal{G}_{1tn})}\right)=O\left(\frac{m_{jn}}{m_{1n}}\right)\to 0 \; \mathrm{as}\; n\to \infty,$$
because $m_{jn}/m_{1n}=o(1)$ for all $j=2,\ldots,J$ as $ n\to \infty$.\\
\underline{\textbf{Assumption C} :} By an easy calculation, we can show that 
$$\frac{\mathbb{E}_0\left(\vert A_{jtn}\vert^{2\tau}\right)}{\sigma_{jtn}^2 \gamma_{jn}^{2\tau-2}}=O(m_{jn}^{2-2\tau})\to 0 \; \mathrm{as}\; n\to \infty.$$  
\begin{lemma}\label{l7}
	Under the assumptions of Theorem~\ref{th2} and for any $\tilde{\theta}$ such that  $\tilde{\theta}-\theta_0=o_p(1)$,
	we have 
	\begin{equation}
	\frac{\partial S_n}{\partial\theta}(\tilde{\theta},g_{\tilde{\theta}})-\frac{\partial S}{\partial\theta}(\theta_0,g_{0})=o_p(1)
	\label{EL71}
	\end{equation}
	and 
	\begin{equation}
	\frac{\partial S_n}{\partial g}(\tilde{\theta},g_{\tilde{\theta}})g_{\tilde{\theta}}^{'}-\frac{\partial S}{\partial g}(\theta_0,g_{0})g_{0}^{'}=o_p(1),
	\label{EL72}
	\end{equation}
	with $\displaystyle g^{'}_{\tilde{\theta}}(.)=\left.\frac{g_\theta}{\partial \theta^T}(.)\right\vert_{\theta=\tilde{\theta}}$.
\end{lemma}
\subsection*{Proof of Lemma~\ref{l7}}
To prove (\ref{EL71}), we need to show that for all $w\in \mathbb{R}^q$ with $\Vert w\Vert =1$, 
$$w^T\left\{\frac{\partial S_n}{\partial\theta}(\tilde{\theta},g_{\tilde{\theta}})-\frac{\partial S}{\partial\theta}(\theta_0,g_{0})\right\}=o_p(1)$$,
which is equivalent to 
\begin{equation}
w^T\left\{\frac{\partial S_n}{\partial\theta}(\tilde{\theta},g_{\tilde{\theta}})-\frac{\partial S_n}{\partial\theta}(\theta_0,g_{0})\right\}=o_p(1)
\label{EqL1}
\end{equation}
and 
\begin{equation}
w^T\left\{\frac{\partial S_n}{\partial\theta}(\theta_0,g_{0})-\frac{\partial S}{\partial\theta}(\theta_0,g_{0})\right\}=o_p(1).
\label{EqL2}
\end{equation}
The proof of (\ref{EqL1}) is similar to that of (\ref{EquiCon}), using the fact that 
$$\sup_{\theta,\, \eta} \left\Vert \frac{\partial^2\tilde{U}_{i}}{\partial \theta\partial \theta^T}(\theta,\eta)\right \Vert \qquad\mathrm{and} \qquad \sup_{\theta,\, \eta} \left\Vert \frac{\partial^2\tilde{U}_{i}}{\partial\theta\partial\eta}(\theta,\eta)\right \Vert$$
are bounded uniformly on $i$ and $n$, and $\tilde{\theta}-\theta_0=o_p(1)$. \\
Now, let us prove (\ref{EqL2}). By the definition of $S(\cdot\, ,\, \cdot)$ (see \ref{DefS}) 
$$\lim_{n\to \infty} \mathbb{E}_0\left(\frac{\partial S_n}{\partial\theta}(\theta_0,g_{0})\right)=\frac{\partial S}{\partial\theta}(\theta_0,g_{0}).$$
Thus, it suffices to prove that 
\begin{equation}
w^T\frac{\partial S_n}{\partial\theta}(\theta_0,g_{0})-w^T\mathbb{E}_0\left(\frac{\partial S_n}{\partial\theta}(\theta_0,g_{0})\right)=o_p(1). 
\label{EqL3}
\end{equation}
Let  
\begin{equation}
w^T\frac{\partial S_n}{\partial\theta}(\theta_0,g_{0})= n^{-1}w^T\xi_{in}\frac{\partial \tilde{U}_{in}}{\partial\theta}(\theta_0,\eta_i^0),
= \Delta_{n1}- \Delta_{n2},
\label{EqL4}
\end{equation}
where 
$$
\Delta_{n1}=n^{-1}\sum_{i=1}^{n}\xi_{in}^{(1)}(\theta_0,\eta_i^0)\left(Y_{in}-\Phi\left(G_{in}(\theta_0,\eta_i^0)\right)\right) \quad
\mathrm{and} \quad 
\Delta_{n2}= n^{-1}\sum_{i=1}^{n}\xi_{in}^{(2)}(\theta_0,\eta_i^0),$$
with  
$$\xi_{in}^{(1)}(\theta_0,\eta_i^0):= w^T\xi_i\Lambda'\left(G_{in}(\theta_0,\eta_i^0)\right)\frac{\partial G_i }{\partial \theta}(\theta_0,\eta_i^0),$$
$$\xi_{in}^{(2)}(\theta_0,\eta_i^0):= w^T\xi_{in}\Lambda\left(G_{in}(\theta_0,\eta_i^0)\right)\phi\left(G_{in}(\theta_0,\eta_i^0)\right)\frac{\partial G_{in} }{\partial \theta}(\theta_0,\eta_i^0),$$
and $\eta_i^0=g_0(Z_{in})$. \\
The proof of (\ref{EqL3}) is then reduced to  proving
\begin{equation}
\mathbb{E}_0\left(\Vert \Delta_{n1}\Vert^2\right)=o(1)\qquad \mathrm{and}\qquad \mathbb{E}_0\left(\Vert \Delta_{n2}-\mathbb{E}_0(\Delta_{n2})\Vert^2\right)=o(1).
\label{EqL5}
\end{equation}
This last part is trivial because $\xi_{in}^{(1)}$ and $\xi_{in}^{(2)}$ are bounded uniformly on $i$ and $n$  (see Assumption A6 and the compactness of $\Theta$, $\mathcal{X}$, and $ \mathcal{Z}$) and by use of the  mixing condition  (\ref{DepH}) and (\ref{DepH2}) in Assumption A3. 
This completes the proof of (\ref{EL71}).   \\

To prove (\ref{EL72}), we remark that
\begin{align}
&\frac{\partial S_n}{\partial g}(\tilde{\theta},g_{\tilde{\theta}})g_{\tilde{\theta}}^{'}-\frac{\partial S}{\partial g}(\theta_0,g_{0})g_{0}^{'}=\nonumber \\&\qquad\qquad\left\{
\frac{\partial S_n}{\partial g}(\tilde{\theta},g_{\tilde{\theta}})-\frac{\partial S}{\partial g}(\theta_0,g_{0})\right\}g_{\tilde{\theta}}^{'}+\frac{\partial S}{\partial g}(\theta_0,g_{0})\left(g_{\tilde{\theta}}^{'}-g_{0}^{'}\right).	
\label{EqL6}
\end{align}
Consider  the second term on the right-hand side in (\ref{EqL6}), where we remark that because $\displaystyle \left\Vert  \frac{\partial S}{\partial g}(\theta_0,g_{0})\right \Vert$ and $\displaystyle\sup_{\theta}\sup_{z}\left\Vert\frac{\partial g_{\theta}(z)}{\partial \theta\partial \theta^T}\right\Vert$ are finite and $\tilde{\theta}-\theta_0=o_p(1)$, 
\begin{equation*}
\frac{\partial S}{\partial g}(\theta_0,g_{0})\left(g_{\tilde{\theta}}^{'}-g_{0}^{'}\right)	=(\tilde{\theta}-\theta_0)\,O\left( \left\Vert  \frac{\partial S}{\partial g}(\theta_0,g_{0})\right \Vert\sup_{\theta}\sup_{z}\left\Vert\frac{\partial g_{\theta}(z)}{\partial \theta\partial \theta^T}\right\Vert\right)=o_p(1).
\end{equation*}

For the first term on  the right-hand side in (\ref{EqL6}), because $g_{\tilde{\theta}}^{'}=O_p(1)$ by Proposition~\ref{prop1}, using similar arguments as when proving (\ref{EL71}) permits one to obtain 
$$
\frac{\partial S_n}{\partial g}(\tilde{\theta},g_{\tilde{\theta}})-\frac{\partial S}{\partial g}(\theta_0,g_{0})=o_p(1).$$
This yields the proof of (\ref{EL72}). $\quad \square$
\begin{lemma} \label{l5}
	Under the assumptions of Theorem~\ref{th2}, we have 
	$$(i) \qquad \left.\frac{d}{d\theta} \frac{\partial Q_{n}}{\partial g}(\theta,g_{\theta})\right\vert_{\theta=\theta_0}(\hat{g}_{0}-g_{0})=o_p(1)$$
	$$(ii) \qquad \left. \frac{\partial Q_{n}}{\partial g}(\theta,g_{\theta})\right\vert_{\theta=\theta_0}(\hat{g}^{'}_{0}-g^{'}_{0})=o_p(1),$$
	where $$\hat{g}^{'}_{0}(.)=\left.\frac{\partial \hat{g}_\theta}{\partial \theta}(.)\right \vert_{\theta=\theta_0} \qquad  \mathrm{and} \qquad g^{'}_{0}(.)=\left.\frac{\partial g_\theta}{\partial \theta}(.)\right \vert_{\theta=\theta_0}.$$
\end{lemma}
\section*{Proof of Lemma \ref{l5}}
To prove  $(i)$, and we note that
\begin{eqnarray*}
	\frac{d }{d \theta}\frac{\partial Q_{n} }{\partial g} (\theta, g_{\theta})&=& 2 \frac{d }{d \theta}\left\{ S_{n}^T (\theta, g_{\theta})M_n  \frac{\partial S_{n} }{\partial g}(\theta, g_{\theta})\right\}\nonumber \\
	&=& 2 \frac{d }{d \theta}S_{n}^T (\theta, g_{\theta})M_n  \frac{\partial S_{n} }{\partial g}(\theta, g_{\theta})+ 2 S_{n}^T (\theta, g_{\theta})M_n  \frac{d }{d \theta}\frac{\partial S_{n} }{\partial g}(\theta, g_{\theta}).
\end{eqnarray*}
One can easily see that 
\begin{equation*}
\frac{d }{d \theta}S_{n} (\theta, g_{\theta})= \frac{\partial S_n}{\partial \theta}(\theta, g_{\theta})+  \frac{\partial S_n}{\partial g}(\theta, g_{\theta})g^{'}_{\theta}
\end{equation*}
and 
\begin{equation*}
\frac{d }{d \theta}\frac{\partial S_n}{\partial g}(\theta, g_{\theta})= \frac{\partial^{2} S_n}{\partial \theta\partial g}(\theta, g_{\theta})+  \frac{\partial^{2} S_n}{\partial g^2}(\theta, g_{\theta})g^{'}_{\theta}.
\end{equation*}
Therefore, we have
\begin{align*}
&\frac{d }{d \theta}\left.\frac{\partial Q_{n} }{\partial g} (\theta, g_{\theta})\right \vert_{\theta=\theta_0} (\hat{g}_0-g_0)= \nonumber\\
& 2 S_{n}^T (\theta_0, g_{0})M_n \left\{ \frac{\partial^{2} S_n}{\partial \theta\partial g}(\theta_0, g_{0})+  \frac{\partial^{2} S_n}{\partial g^2}(\theta_0, g_{0})g^{'}_{0}\right\} (\hat{g}_0-g_0) \nonumber\\
&\qquad+2\frac{\partial S_n}{\partial g}(\theta_0, g_{0})M_n\left\{\frac{\partial S_n}{\partial \theta}(\theta_0, g_{0})+  \frac{\partial S_n}{\partial g}(\theta_0, g_{0})g^{'}_{\theta}\right\}(\hat{g}_0-g_0).
\end{align*}
By Lemma~(\ref{l1}) and  $S(\theta_0,g_0)=0$, we obtain
\begin{equation}
S_n(\theta_0,g_0)=S_n(\theta_0,g_0)-S(\theta_0,g_0)=o_p(1).
\label{EQ0}
\end{equation}
In addition,  we have
\begin{eqnarray}
\left \Vert \frac{\partial^{2} S_n}{\partial \theta\partial g}(\theta_0, g_{0})(\hat{g}_0-g_0)\right \Vert&=&n^{-1}\left \Vert\sum \xi_{in} \frac{\partial^{2} \tilde{U}_{in}}{\partial \theta\partial \eta}(\theta_0, \eta_i)(\hat{g}_0(Z_{in})-g_0(Z_{in}))\right \Vert \nonumber \\ 
&\leq & n^{-1}\sum \sup_{i,n} \Vert\xi_{in}\Vert \sup_{\eta}\left \Vert \frac{\partial^{2} \tilde{U}_{in}}{\partial \theta\partial \eta}(\theta_0, \eta)\right \Vert \Vert\hat{g}_0-g_0\Vert \nonumber\\
&=&  o_p(1),
\label{EQ3}
\end{eqnarray}  
because $\xi_i$ is bounded uniformly on $i,\, n$ and $\theta$ (Assumption A6), $\Vert\hat{g}_0-g_0\Vert=o_p(1)$ by  Proposition~\ref{prop1}, and 
$$\sup_{i,\, n}\sup_{ \eta}\left \Vert \frac{\partial^{2} U_{in}}{\partial \theta\partial \eta}(\theta_0, \eta)\right \Vert< \infty.$$
Using similar arguments as in the proof of (\ref{EQ3}), we obtain  
\begin{eqnarray}
\left \Vert \frac{\partial^{2} S_n}{\partial g^2}(\theta_0, g_{0})(\hat{g}_0-g_0)g^{'}_{0}\right \Vert&=&n^{-1}\left \Vert\sum \xi_i \frac{\partial^{2} U_{in}}{\partial \eta^2}(\theta_0, \eta_i)(\hat{g}_0(Z_{in})-g_0(Z_{in}))g^{'}_0(Z_{in})\right \Vert\nonumber \\
&= & o_p(1),
\label{EQ4}
\end{eqnarray}

\begin{eqnarray}
\left \Vert \frac{\partial S_n}{\partial g}(\theta_0, g_{0})(\hat{g}_0-g_0)g^{'}_{0}\right \Vert&=&n^{-1}\left \Vert\sum \xi_{in} \frac{\partial U_{in}}{\partial \eta}(\theta_0, \eta_i)(\hat{g}_0(Z_{in})-g_0(Z_{in}))g^{'}_0(Z_{in})\right \Vert\nonumber \\
&= & o_p(1),
\label{EQ5}
\end{eqnarray}
and 
\begin{eqnarray}
\left \Vert \frac{\partial S_n}{\partial \theta}(\theta_0, g_{0})(\hat{g}_0-g_0)\right \Vert&=&n^{-1}\left \Vert\sum \xi_{in} \frac{\partial U_{in}}{\partial \theta}(\theta_0, \eta_i)(\hat{g}_0(Z_{in})-g_0(Z_{in}))\right \Vert\nonumber \\
&= & o_p(1).
\label{EQ6}
\end{eqnarray}
Combining (\ref{EQ0})-(\ref{EQ6}) with Assumption A10 permits one to have
$$\frac{d }{d \theta}\left.\frac{\partial Q_{n} }{\partial g} (\theta, g_{\theta})\right \vert_{\theta=\theta_0} (\hat{g}_0-g_0)=o_p(1).$$
This yields the proof of $(i)$.\\
The proof of $(ii)$ follows along similar lines as (i) and hence is omitted. $\quad \square$
\begin{lemma}\label{l6} 
	Under the assumptions of Theorem~\ref{th2}, we have 
	$$S_{n}(\theta,\hat{g}_\theta)-S_{n}(\theta,g_\theta)=r^{(1)}_{n}(\theta),$$  where  
	$$\sup_{\theta}\left\Vert \frac{\partial
	}{\partial\theta}r_{n}^{(1)}(\theta) \right\Vert=o_p(1), \qquad \mathrm{and}\qquad\sup_{\theta}\left\Vert \frac{\partial^2}{\partial\theta \partial\theta^T}r_{n}^{(1)}(\theta) \right\Vert=o_p(1)$$
\end{lemma}

\subsection*{Proof of Lemma \ref{l6}}
By applying Taylor's theorem  to $\tilde{U}_{i}(\theta, \cdot)$ for each $\theta\in \Theta$, we obtain 
\begin{eqnarray*}
	S_{n}(\theta,\hat{g}_{\theta})-S_{n}(\theta,g_{\theta})&=&n^{-1}\sum_{i=1}^{n}\xi_{in}\left(\tilde{U}_{in}(\theta,\hat{g}_{\theta})-\tilde{U}_{in}(\theta,g_{\theta})\right)\\
	&=&n^{-1}\sum_{i=1}^{n}\xi_{in}\left(\hat{g}_{\theta}(Z_{in})-g_{\theta}(Z_{in})\right)\\
	&& \qquad \times \int_{0}^{1}\frac{\partial\tilde{U}_{in}}{\partial\eta}\left(\theta , g_{\theta}(Z_{in})+t\left(\hat{g}_{\theta}(Z_{in})-g_{\theta}(Z_{in})\right)\right)dt\\
	&:=&r^{(1)}_n(\theta).
\end{eqnarray*}
Because the instrumental variables are bounded uniformly on  $i,\, n ,$ and $\theta$ (Assumption A6),  
$\displaystyle \sup_{\theta\in \Theta}\left \Vert \hat{g}_{\theta}-g_{\theta}\right\Vert $, $\displaystyle\sup_{\theta\in \Theta}\max_{j=1,\ldots,p+1}\left \Vert \frac{\partial}{\partial \theta_j}\left(\hat{g}_{\theta}-g_{\theta}\right)\right\Vert $ and  $\displaystyle\sup_{\theta\in \Theta}\max_{1\leq i ,j\leq p+1}\left \Vert \frac{\partial^{2}}{\partial\theta_i\partial \theta_j}\left(\hat{g}_{\theta}-g_{\theta}\right)\right\Vert $ are all of order $o_p(1)$ by Proposition~\ref{prop1},  
it suffices to show that
\begin{equation}
\sup_{\theta,\eta}\sup_{i}\left\Vert \frac{\partial\tilde{U}_{in}}{\partial\eta}\left(\theta , \eta\right)\right\Vert =O_{p}(1)
\label{EQl51}
\end{equation}
\begin{equation}
\sup_{\theta,\eta}\sup_{i}\left\Vert \frac{\partial}{\partial\theta}\frac{\partial\tilde{U}_{in}}{\partial\eta}\left(\theta , \eta\right)\right\Vert =O_{p}(1)\qquad \mathrm{and}\qquad \sup_{\theta,\eta}\sup_{i}\left\Vert \frac{d^{2}}{\partial \theta\partial\theta^{T}}\frac{\partial\tilde{U}_{in}}{\partial\eta}\left(\theta , \eta\right)\right\Vert =O_{p}(1).
\label{EQl52}
\end{equation}
Equation (\ref{EQl51}) is already proved in the proof of Lemma~\ref{l2} (see (\ref{U2Borne})). The
proof of (\ref{EQl52}) can be established in a similar manner and is thus omitted.$\quad \square$

\bibliographystyle{model5-names}
\bibliography{biblio}

\end{document}